# Quantum Algorithms for Molecular Response Functions: Resolvent Kernels, Time-Domain Sampling, Query Lower Bounds with Matched Single-Point Calibers, and a Six-Route Comparison

**Yunlong Liu,[1] Yujin Tang,[1] Miao Xu,[2] and Qing-Xing Xie[1]**

[1]School of Physics, Hubei University, Wuhan 430062, China

[2]Normal College, Hubei University, Wuhan 430062, China

Corresponding author: Qing-Xing Xie (xieqx@hubu.edu.cn).

## Abstract

Quantum algorithms for molecular response functions have mostly been studied one route at a time, most prominently the indicator-binning algorithm of Kharazi et al. We compare six routes within one accounting framework and state the conversions between their resource calibers. On the resolvent route the Lorentz-broadened spectrum is exactly a resolvent expectation; at finite $\gamma$ two boundary terms, a reversal term and a permanent-dipole pseudo-line, enter a disjoint error budget. Because the kernel subnormalization is exactly 1, the published binning accounting carries a removable $\Theta(\alpha + 1 + \omega)$ factor (49.69 × and 7.75 × at the definitional caliber), a correction that applies to both routes. Estimating the square root of the kernel amplitude lowers the readout cost from $\beta^2/\tilde{\mu}_2$ to $\beta/\sqrt{\tilde{\mu}_2}$. On the time-domain route we prove a single-point amplitude-estimation theorem and convert the generalized QPE of Loaiza et al. to the same accounting. The lower bounds are $\widetilde{\Theta}(1/\gamma)$ for parameter location, $\Theta(1/(\gamma\varepsilon))$ for windowed values, $\Omega(n/\gamma)$ inside a restricted correlation-sampling class, and $\Omega(1/(\varepsilon\gamma^2))$ on the Lorentzian object; in the single-circuit-resolution caliber, $\gamma^{-n}$ is a grid-coverage artifact. Six routes, their calibers, an $n = 2$ assembly with a component-grade coherence, and a phase diagram over the tested ensembles form one accounting map.

## 1 Introduction

### 1.1 Background and problem

The $n$-th order (frequency-domain) response function $\alpha^{(n)}(\omega_n, \ldots, \omega_1)$ of a molecule governs its linear and nonlinear optical properties; its sum-over-states (SOS) form involves transition dipoles $d_{n0} = \langle\psi_0|D|\psi_n\rangle$ and excitation energies $\omega_{n0}$ of the electronic Hamiltonian $H_0$ (notation in Sec. 3; atomic units throughout). On fault-tolerant quantum computers, the reference algorithm family is the excite–filter–read construction of Kharazi et al. [1]: prepare the ground state (assumption A1), apply a dipole operator $D$ through its block encoding $U_D$ (normalization $\beta$), apply a QSP polynomial approximation of a spectral *indicator function* $\mathbb{1}_{[a,b]}(H_0)$ (block encoding of $H_0$ with normalization $\alpha$), apply $D'$, and read the expectation through a Hadamard test with amplitude estimation (AE). The measured object is a *bin intensity*; the response is assembled from bins via adaptive bisection, giving the headline complexity $\tilde{O}(\alpha^{2n}\beta^{n+1}/(\gamma^n\varepsilon))$ for the $n$-th order response (their Eq. 33), with $\gamma$ the spectral resolution and $\varepsilon$ the amplitude precision.

Three questions motivate this paper.

1. **Exact lineshapes at binning prices?** The indicator window introduces three approximation layers (bin merging, leakage outside the window, center-frequency approximation). The physical object — a Lorentz-broadened spectrum at phenomenological width $\gamma$ — is a *convolution* of the line spectrum with a Lorentzian kernel. Can a quantum algorithm sample that kernel directly, at what cost, and what boundary terms appear at finite $\gamma$?
2. **Is $\gamma^{-n}$ necessary?** The $\gamma^{-n}$ factor of Eq. 33 is heuristically justified only for a single frequency variable; whether it is an information-theoretic requirement or an artifact of $n$-dimensional grid coverage has been open (Kharazi et al. themselves flag this; see also [10] for upper bounds only).
3. **Which sampling arm for which spectrum?** Indicator binning, smooth-kernel sampling, and time-domain correlation sampling are three measurement models for the same object; sparse-recovery and super-resolution theory suggest their failure modes differ qualitatively. No comparison with explicit caliber conversions, and no empirical phase diagram, existed in the literature we surveyed (search window and disclaimers in Sec. 2); we supply cross-caliber maps with the conversions stated and same-caliber rows where available.

## 1.2 Contributions

1. **Resolvent route: exact lineshapes, boundary terms, and a tight accounting (Secs. 4.1–4.4).** Starting from the standard identity that $\mathrm{Im}\langle\psi_0|D(H_0-\omega-i\gamma)^{-1}D|\psi_0\rangle$ *is* the Lorentz-broadened spectrum, we quantify its two finite-$\gamma$ boundary terms — a reversal term of relative size $(\gamma/2\omega^*)^2 C_{\mathrm{sys}}$ ($C_{\mathrm{sys}} = 2.05$ $H_2O$-z, 1.16 LiH-z) and an $n = 0$ permanent-dipole pseudo-line ($|d_{00}|^2 = 11.01$ for LiH-z) — in a mutually disjoint budget with zero- or $\times$ 2-cost remedies. The kernel block encoding carries subnormalization exactly 1, so the published accounting of Kharazi et al. contains a removable $\Theta(\alpha+1+\omega)$ factor (definitional ratio 49.7 $\times$ $H_2O$-z, 7.75 $\times$ LiH-z): under the tight reading the two routes share their per-point order, and the resolvent route's value lies in the error structure and in a frequency-independent preparation circuit (saving $(\beta/\xi)^2$), with a square-root-kernel readout improving the signal layer to $\beta/\sqrt{\tilde{\mu}_2}$.
2. **Nested response and the $n = 2$ assembly (Sec. 4.5, SM §B).** Theorem 4 gives $\tilde{O}((d_1 + d_2)(\zeta^{(2)}/\tilde{\mu}^{(2)})\varepsilon_{\mathrm{rel}}^{-1})$ with $\zeta^{(2)} = \beta^3$ for the nested kernel expectation; an assembly lemma maps each Orr–Ward permutation term [33, 34] to one such expectation, and a new proposition carries the physical $\beta_{ijk}$ error bill through a component-grade coherence $\kappa_{\mathrm{asm}}$ (machine-precision bookkeeping checks in SM §B.5).
3. **Kernel trade-off and the time-domain route (Secs. 4.6–5).** At matched resolution no kernel strictly dominates (degrees $576 < 1202 < 1376 \ll 2552$ at $\tilde{\rho} = 0.02$, $\varepsilon = 10^{-6}$; leakage beyond $4\rho$ 6.4% vs 0; shape bias 0 vs ~ 14% vs ~ 33%), which yields three selection criteria. A single-point amplitude-estimation theorem, the same-accounting conversion of GQPE [10] (fixed point $(\pi/2)n\beta^{n+1}/(\gamma\varepsilon)$, no $\alpha$ in the additive-precision reading), and a transfer of classical conditional ESPRIT stability to the quantum sampling-cost model complete the route.
4. **Query lower bounds (Sec. 6).** $\Omega(1/\gamma)$ for parameter location (a parallel regime on a toy family, *not* $n/\gamma$); $\Theta(1/(\gamma\varepsilon))$ for windowed value tasks; $\Omega(n/\gamma)$ multi-variable depth within the restricted direct correlation-sampling class; $\Omega(1/\varepsilon)$ readout; and $\Omega(1/(\varepsilon\gamma^2))$ on the Lorentzian object ($s^* = \Theta(1)$ families), with exponent-matched upper bounds across the stated oracle models. Consequently $\gamma^{-n}$ is a grid-coverage artifact in the single-circuit-resolution caliber.

5. **Cross-route accounting and empirics (Secs. 7–8).** Table 14a compares per-route normalized single-point costs and Table 14b documents the native semantics with the conversions; the $\varepsilon^2$-caliber mechanism is carried by the table. On the tested ensembles kernel sampling is most accurate on few-line and continuous spectra (median $L^2$ $3 \times 10^{-4}$–$4 \times 10^{-3}$, vs bin 0.06–0.23 and ESPRIT 0.20–0.79), while the time-domain arm localizes sharp off-grid lines grid-independently ($\mathrm{err}/\varepsilon_q \approx 0.45$; 2D support 20/20).

Numerical claims are extracted mechanically from the archived result files, exceptions being flagged at their point of use (SM §H).

---

## 2 Related work

**Indicator-binning family and precursors.** Cai et al. [2] solve the linear system $(H - \omega - i\gamma)|x\rangle = D|\psi_0\rangle$ by HHL-type algorithms (modernized estimate $\tilde{O}(N^7 N_e/(\gamma\varepsilon^2))$ — ours; Cai et al.'s original form $O(\log(D)s^2\kappa^2/\varepsilon)$ carries no electron-count factor, the $N^7 N_e$ entering through our block-encoding modernization; the $\varepsilon^2$ mechanism is discussed in Sec. 7.2); Kosugi–Matsushita [3] estimate response matrix elements one at a time (success probability can vanish by symmetry); Huang et al. [4] is variational (no a priori runtime); Roggero–Carlson [5] gave a dynamic linear-response quantum algorithm for nuclear physics; Ding et al. [6] use Gaussian-filtered QPE for multi-eigenvalue search (their Gaussian-filter choice receives independent support from our kernel trade-off, Sec. 4.6); Bruschi et al. [7] simulate 2D electronic spectroscopy digitally without a query-complexity theory.

**2025–2026 time-domain competitors** (calibers fixed by the literature survey underlying Sec. 2, search date 2026-10-19 — protocol in SM §I; all novelty statements below are scoped to that survey):

- *Wang, Xiong, Cai, Yuan* [8] (PRL 135, 230602): $n$-time correlation functions without ancilla qubits. Published version (v3): finite parameter-shift rule ($\tau_- = \pi/4$, $\tau_+ = 1/2\log(1+\sqrt{2})$, $2^{n-1}$ shift points) with Trotterized QITE ($n_+ \leq n-1$ applications) or mid-circuit measurement; bang-bang pulse approximation with total error $O(n/\Lambda)$. Precision is shot-noise $\varepsilon^{-2}$; no amplitude estimation; no $\gamma$-/resolution-aware query theorem. NISQ positioning (SSH, Schwinger, OTOC demonstrations).
- *Xiong, Wang, Cai, Yuan* [9] (arXiv:2604.16164): unified generalized-parameter-shift framework expressing the $m$-th order response as mixed derivatives in the pump amplitude; Trotterized real-time evolution plus expectation measurements; per-time-point circuit count independent of order. No query-complexity or sampling-theory theorem.
- *Loaiza et al.* [10] (Quantum 9, 1822): generalized quantum phase estimation (GQPE) for nonlinear spectroscopy, with Theorems 1–2 (fixed-point $\tilde{O}(D\Omega/(\eta\varepsilon))$ $e^{\pm i\hat{H}}$ queries *with* amplitude estimation; full-spectrum output-distribution sampling with $O(\Omega^4/\varepsilon^4)$ samples and no grid factor) and a single-ancilla variant (SM §C, Eq. C9–C10). This occupies the fault-tolerant-plus-theorems end of the time-domain route; Sec. 5.2 converts it to our caliber, and their upper bounds enter our matching analysis (Sec. 6) on the upper-bound side.

- *Ono* [11] (PRL 135, 026401): a *classical* framework extracting nonlinear response functions from real-time evolution via functional derivatives (tensor networks); a classical counterpart, not a competitor.
- *Steffens et al.* [12] (NJP 19, 033005): quantum matrix-pencil/ESPRIT-family spectral estimation for damped sums of sinusoids — the closest precedent to our "quantum sampling + classical sparse reconstruction" architecture; single-variable, with state preparation outside its cost model.

**Classical toolbox.** Super-resolution and off-grid compressed sensing: Candès–Fernandez-Granda [27, 28]; Tang–Bhaskar–Shah–Recht [29]; atomic-norm denoising: Bhaskar–Tang–Recht [30] (cited with the conditions listed in SM §F); sparse FFT: [31, 32]. Deconvolution with supersmooth kernels: Fan [35]. ESPRIT/matrix-pencil background: [36, 12]. Multi-parameter quantum metrology (used only as *machinery*, with the variance-vs-query-caliber caveat of [23]): [24, 25, 26]; AE optimality: [21, 22].

**Positioning (scoped to the surveyed literature, search date 2026-10-19).** The fault-tolerant time-domain-plus-theorems end is occupied by [10]; the NISQ end by [8, 9]. What remained open — and what this paper contributes — are: (a) exact-lineshape (resolvent) algorithms with a tight-caliber cost theorem and an error budget including both finite-$\gamma$ boundaries; (b) *lower* bounds ($\gamma^{-1}$ necessity; the parallel-regime observation that $\Omega(n/\gamma)$ does not hold on the toy family; multi-variable depth $\Omega(n/\gamma)$); (c) a sparse-prior algorithm at the same caliber (the sparse-recovery arm of Sec. 5.3) — its benefit is a *target to be tested* against output-distribution sampling, which we do numerically on the ESPRIT side and analytically on the ANM side; (d) the cross-caliber six-route table (with proved/extrapolated status separated); and (e) the empirical phase diagram. We claim no priority on the circuit primitives themselves (Hadamard-test correlation estimation predates us [5], as does GQPE [10]).

---

## 3 Preliminaries and assumptions

**Notation** (self-contained table below; atomic units throughout).

| symbol | meaning |
|---|---|
| $N$, $N_e$ | spin-orbital basis functions (= qubit count under Jordan–Wigner); electron count |
| $H_0$, $(\lambda_j, \psi_j)$ | electronic Hamiltonian (Born–Oppenheimer, fixed nuclei); eigenpairs, shifted so $\lambda_0 = 0$ |
| $\omega_j = \lambda_j$; $G = \lambda_1$ | excitation energies (after the shift); spectral gap |
| $D$ | dipole operator |
| $d_{mn}$, $s_n$ | matrix elements $\langle\psi_m\|D\|\psi_n\rangle$; line strengths $\|d_{n0}\|^2$ |
| $\alpha^{(n)}$, $\chi^{(n)}$ | $n$-th order frequency-domain / time-domain response |
| $\gamma$ | Lorentz broadening (HWHM) = target spectral resolution |

| symbol | meaning |
|---|---|
| $W = \omega_{\max}$ | spectral width; two encodings: centered $\alpha = W/2$ (GQPE tight fit), uncentered $\alpha \geq W$ (Theorem 1's shifted LCU; the numerics use $\alpha_{\mathrm{ED}} = W$) |
| $\bar{S}_\gamma := \gamma\, \mathrm{Im}R$, $S_\gamma^{\mathrm{phys}} := \mathrm{Im}R$ | the QSP even-kernel expectation object ($\bar{S}_\gamma = \tilde{\mu}_2/c$, peak $\approx s^*$) vs the physical broadened spectrum ($\bar{S}_\gamma = \gamma S_\gamma^{\mathrm{phys}}$, peak $\approx s^*/\gamma$) — conversion dictionary in Sec. 4.1 |
| $L$ | window logarithm $\ln(4A_1/(\gamma A_s \varepsilon))$ in instance family F2; $\ln(4A_1/(s^*\varepsilon))$ ($\gamma$-free) in family F1 (Secs. 5.1, 7.3, D.2–D.3; distinct from the physical response $L(\omega)$ of Sec. 4.1) |
| $U_H$: $(\alpha, m, 0)$-BE; $U_D$: $(\beta, k, 0)$-BE | block encodings of $H_0$ resp. $D$ (qubitization sense [15]); normalizations $\alpha$, $\beta$ |
| $\xi = \lVert D \lVert \psi_0 \rangle \rVert$ | dipole signal amplitude on the ground state |
| $\zeta$ | composite subnormalization of a sandwiched circuit |
| $\mathbb{1}_{[a,b]}$, $\tilde{\mathbb{1}}$ | spectral indicator and its polynomial approximation |
| $\varepsilon$, $\varepsilon_{\mathrm{rel}}$, $\tilde{O}(\cdot)$ | additive precision; relative precision; $\tilde{O}$ suppresses logarithms |

(Symbol overloads retained for continuity, each disambiguated at its site: the noise $\eta_j$ and the GQPE resolution $\eta_L$ vs $N_e$; the window logarithm $L$ vs $L(\omega)$ vs the $L^2$ error; the sparsity $s$ vs line strengths $s_n$; the dipole $D$ vs the GQPE order $D$; $W$ vs $W_{\mathrm{tot}}$; the normalization $\beta$ vs the hyperpolarizability components $\beta_{ijk}$; $\alpha$ vs the response functions $\alpha^{(n)}$.)

Complexities are stated as $U_H$ query counts throughout, with dipole ($U_D$) queries listed separately — never mixed. **Resource models (calibers; the term is our shorthand for the cost metric and accounting convention).** A *resource caliber* specifies which cost is counted: total evolution time (sum of all oracle durations over the full execution), circuit depth (serial evolution time within one circuit), $U_H$/dipole query counts, circuit copies, or total samples; formal definitions in SM §E.1. Comparisons across different calibers are marked as such: Table 14a is a per-route normalized cost map (each route at relative-to-peak precision on its own measured object), and Table 14b is a native-semantics reference map with the conversions stated (entries there are not mutually comparable).

Tool-chain pointers: QSP simulation of the evolution segments follows [14]; eigenstate-filtering ground-state preparation [16] enters only through assumption A1; fast matrix inversion / Green's-function machinery [17] underlies the route-(ii) alternative of Sec. 4.2; QSP phase factors are computed with the pyqsp toolchain (QSPPACK family [18]); oblivious/robust amplitude amplification [19] is available where preparation boosting is needed; the HHL ancestor of route ③ is [20].

**Assumptions** (project-wide set, cited by number). A1 (exact ground state, as in Kharazi et al. [1]; its relaxation is left to future work); A2 (electronic degrees of freedom only — nuclei fixed); A3 (semiclassical light field, electric-dipole coupling); A4 (zero temperature, pure ground-state ensemble); A5 (block-encoding subnormalization controls success probability — our Lemma Q *tightens* the relevant normalization); A6 (peak-like sparse spectra — attacked here by the sparse-recovery arm and the continuous-spectrum section); uniform phenomenological $\gamma$ (its microscopic derivation is left to future work).

**Lemma Q (QET tight caliber).** Let $A$ be Hermitian with an $(\alpha, m, 0)$-block encoding $U_A$, and let degree-$d$ QSP phases realize a polynomial $P$ with $\|P\|_{[-1,1]} \leq 1$ and Re$P = \tilde{P}$. In the projected form (GSLW Thm 17) the interleaved sequence gives $\langle 0^m|U_\Phi|0^m\rangle = P(A/\alpha)$, and in the real-polynomial $|+\rangle$ packaging (GSLW Cor 8/18) $\langle 0^{m+1}|U_\Phi|0^{m+1}\rangle = \tilde{P}(A/\alpha)$: either way a block encoding with subnormalization **exactly 1** [13] (theorem numbers follow the arXiv version's numbering; the STOC'19 proceedings version numbers differ). Consequently the composite $U_{D'}\, U_\Phi\, U_D$ (independent ancillas) is a $(\beta\beta', \cdot, 0)$-block encoding of $D'\tilde{P}(A/\alpha)D$ — subnormalization $\zeta = \beta\beta'$, *without* any $(\alpha + \omega)$ factor. The published accounting of Kharazi et al. carries $\zeta_\omega = (\alpha + 1 + \omega)\beta^2$ (their Eq. 12) — valid but conservative (any (1)-BE is also an $(s)$-BE for $s \geq 1$); the origin of the "+1" is not explicit in the parts of the paper and SI we read, and we retain both calibers side by side. This factor enters their Theorem 2 headline multiplicatively with the degree factor (Sec. 4.4). Matrix-level demonstrations (composite full-zero block equal to $D(P + iQ)(A/s)D/\beta^2$ to $1.67 \times 10^{-15}$; shared-ancilla violation $9.48 \times 10^{-1}$; three BE-form/interleaving conventions mutually consistent to $\leq 2.3 \times 10^{-15}$, one $O(1)$ mixing pitfall identified) are archived in the public data package (checks K3/K4, SM §H).

**Amplitude estimation.** Brassard et al. Thm 12 [21]; the $\sqrt{p(1-p)} \leq 1/2$ argument covers all success probabilities; optimality by [22].

---

# 4 The resolvent route: exact lineshapes, tight calibers, kernel trade-offs

## 4.1 Resolvent identities and the two finite-$\gamma$ boundaries

Shift the spectrum so $\lambda_0 = 0$ and insert the resolution of identity in the eigenbasis (all $n$, including $n = 0$):

$$R(\omega) \equiv \langle\psi_0|D(H_0 - \omega - i\gamma)^{-1}D|\psi_0\rangle = \sum_n \frac{s_n}{\omega_{n0} - \omega - i\gamma} \qquad (1)$$

$$R_n(\omega_n, \dots, \omega_1) = \sum_{j_1,\dots,j_n} \frac{d_{0j_n} \cdots d_{j_1 0}}{\prod_{k=1}^n (\omega_{j_k 0} - \omega_k - i\gamma)} \qquad (2)$$

Two structural facts follow. First, Im$R(\omega) = \pi(P_\gamma * S)(\omega)$ with $P_\gamma(x) = \gamma/\pi(x^2 + \gamma^2)$ and $S = \sum_n s_n\, \delta$ the Dirac line measure: **the imaginary part of the resolvent expectation is exactly the Lorentz-broadened line-strength spectrum** — the physical object itself, with no bin-merging, no leakage, no center-frequency approximation. **Conversion dictionary (used throughout).** The QSP even kernel of Sec. 4.2 measures $\bar{S}_\gamma := \gamma\,\text{Im}R = \tilde{\mu}_2/c$ (peak $\approx s^*$ — no $1/\gamma$ enhancement, which sits in the classical rescaling), while the physical broadened spectrum is $S_\gamma^{\text{phys}} = \text{Im}R = \bar{S}_\gamma/\gamma$ (peak $\approx s^*/\gamma$); every cost statement quoting $\tilde{\mu}_2$ or $s^*$ is in $\bar{S}_\gamma$ units, and additive precision $\varepsilon$ on $S_\gamma^{\text{phys}}$ corresponds to relative precision $\varepsilon\gamma/s^*$ on $\bar{S}_\gamma$ at the peak (this conversion drives the Lorentzian-object corollary of Sec. 6). Second, the physical absorptive response $\mathrm{Im}\alpha^{(1)}(\omega) = L(\omega) - L(-\omega)$ differs from $\mathrm{Im}R$ by the **reversal term** $L(-\omega)$, and Eq. (1) contains the $n = 0$ **pseudo-line** $|d_{00}|^2\gamma/(\omega^2 + \gamma^2)$ (width $\gamma$ at $\omega \approx 0$; absent from the physical response by odd-part cancellation). At a dominant line $\omega^*$, the reversal ratio over the *physical* $n \geq 1$ lines is $(\gamma/2\omega^*)^2 C_{\mathrm{sys}}$, with $C_{\mathrm{sys}} = (4\omega^{*2}/s^*)\sum_{n\geq1} s_n/(\omega_n + \omega^*)^2$. This is a system constant, and it has no universal small bound. With the $\lambda_0 = 0$ shift all $\omega_n \geq 0$, so the denominators are bounded below by $\omega^*$ and $C_{\mathrm{sys}} \leq 4\sum_{n\geq1} s_n/s^*$; a large value arises when the total oscillator strength greatly exceeds the strongest line, or when $s^*$ is small. Measured values are 2.05 for $H_2O$-z and 1.16 for LiH-z. (The $n = 0$ term is excluded from $C_{\mathrm{sys}}$: it *is* the pseudo-line of the next paragraph, removable analytically at zero $U_H$ cost, and counting it here as well would double-charge one physical contribution — for LiH-z the $n = 0$ term contributes 98.6% of the raw reversal numerator, which is why the earlier all-$n$ sum read 82.67.) Their *relative shares* at a fixed positive transition frequency vanish quadratically in $\gamma$ (two-point log-log slopes 1.996–2.000 on the archived scan rows; the pseudo-line's *absolute* magnitude $|d_{00}|^2\gamma/(\omega^2 + \gamma^2)$ vanishes only linearly at fixed $\omega > 0$ — all quadratic statements are share-level), and both have cheap remedies (Sec. 4.3).

The **time-domain bridge** $(H_0 - \omega - i\gamma)^{-1} = i\int_0^\infty e^{-\gamma s}\, e^{i(\omega - H_0)s}\, ds$ makes $R(\omega)$ the one-sided Fourier–Laplace transform of the damped correlation function $C_2(s) = \langle\psi_0|De^{-iH_0 s}D|\psi_0\rangle$ — the resolvent route (this section) and the time-domain route (Sec. 5) are two sampling models of one object, and the $\gamma^{-1}$ degree factor below is the frequency-domain face of the correlation time $\Theta(1/\gamma)$.

**Machine-precision verification** (Table 1): dense complex linear solves vs. spectral sums agree to $\leq 2.2 \times 10^{-12}$ ($n = 1$, three systems, $\omega$-grid covering two strongest lines, the inter-peak region, and 0); nested $n = 2$ objects to $\leq 2.0 \times 10^{-12}$ at near-resonant frequency pairs; the time bridge to $\leq 1.3 \times 10^{-6}$ (integration-error dominated). $H_2O$-x with three parallel $D_x$ slots is *exactly zero* by symmetry (reflection operator argument; archived residual $|R_2| \sim 10^{-10}$) — a symmetric-zero phenomenon on the nested-resolvent object, entering Theorem 4's SNR condition (the parity criterion of SM §B.1).

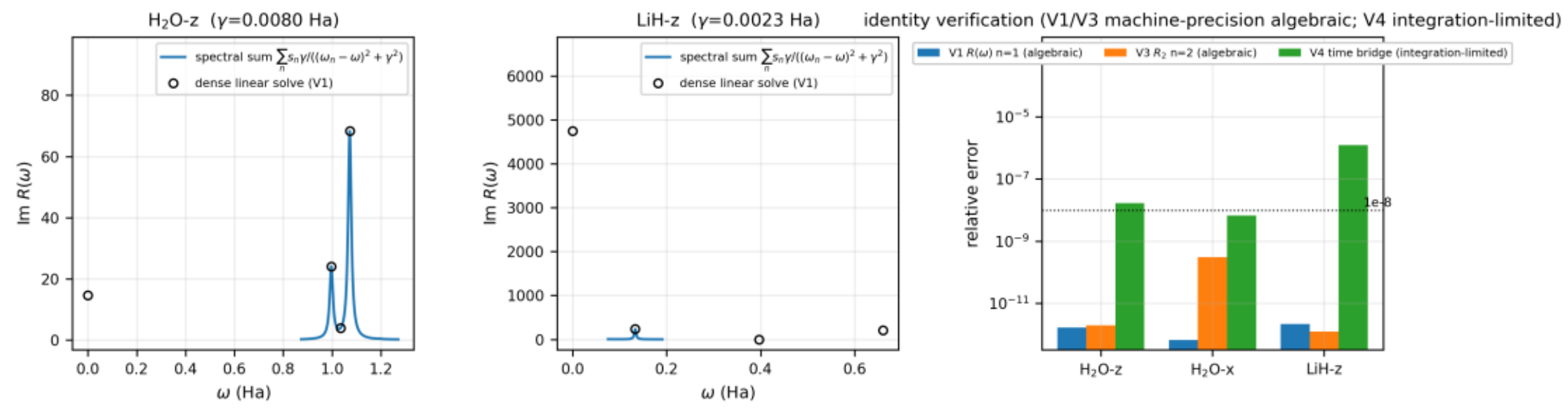

*Figure 1. Resolvent identity. (a,b)* $\mathrm{Im}R(\omega)$ *from the archived full line lists (sto-3g ED,* $H_2O$*-z and LiH-z, archived* $\gamma$*) with the dense-linear-solve markers of the V1 verification overlaid. (c) Verification summary: V1/V3 are machine-precision algebraic checks (dense solve vs spectral sum), V4 is an integration-limited time-bridge agreement (*$6.7\times10^{-9}$*–*$1.2\times10^{-6}$*, trapezoid-integration error dominated);* $H_2O$*-x V3 shows* $|R_2| = 3.10\times10^{-10}$*, the symmetry-zero residual.*

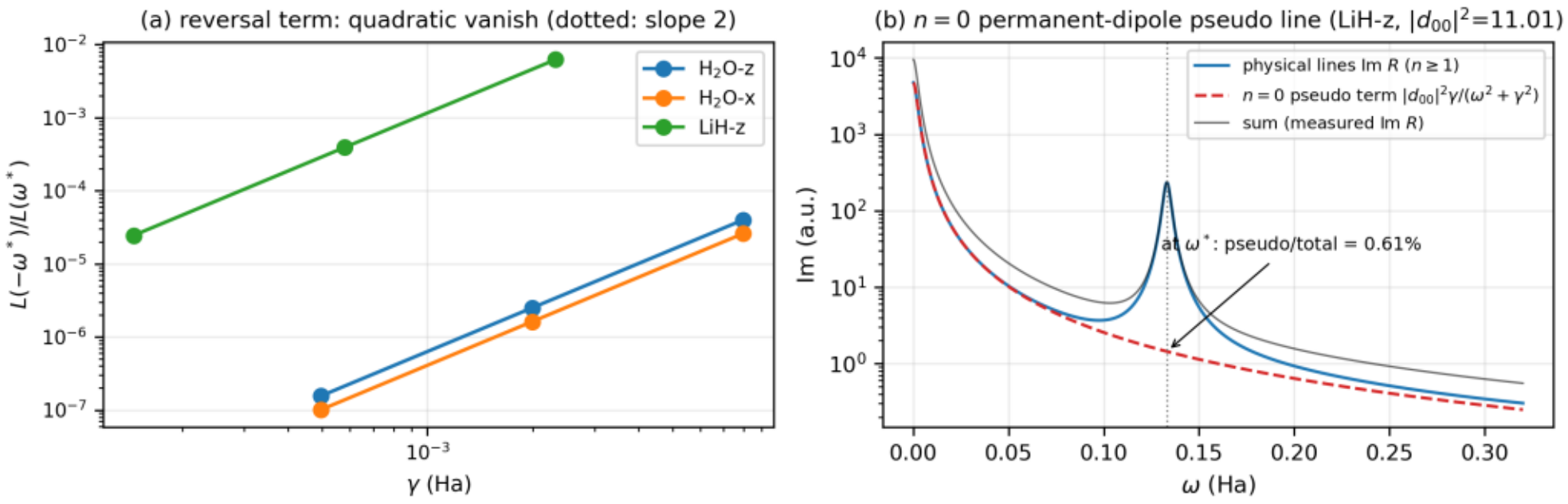

*Figure 2. The two finite-*$\gamma$ *boundaries. (a) Reversal ratio* $L(-\omega^*)/L(\omega^*)$ *vs* $\gamma$ *(log-log) for the three systems with slope-2 guides. (b) The* $n=0$ *permanent-dipole pseudo-line of LiH-z, with* $|d_{00}|^2 = 11.01$*, vs the physical (*$n \geq 1$*) spectrum and their sum; at* $\omega^*$ *the pseudo share is* $0.61\%$ *(the* $n \geq 1$ *reversal share there is* $0.009\%$ *— Table 3). All panels from the archived numerical results.*

**Table 1. Identity verification summary** (all energies and widths in Ha; the V3 column is a relative error except for $H_2O$-x, whose nested object is an exact symmetry zero — there the entry is the absolute residual $|R_2|$).

| system | dim | $N_e$ | $\gamma$ | $\omega^*$ | $s^*$ | V1 max rel err | V3 rel err / abs | V4 rel err |
|---|---|---|---|---|---|---|---|---|
| $H_2O$-z | 441 | 10 | 0.007953889 | 1.0728 | 0.5411 | 1.64e-12 | 1.97e-12 | 1.69e-08 |
| $H_2O$-x | 441 | 10 | 0.007953889 | 0.8259 | 1.8154 | 6.58e-13 | 3.10e-10 (sym.) | 6.66e-09 |
| LiH-z | 225 | 4 | 0.002318753 | 0.1331 | 0.5403 | 2.16e-12 | 1.24e-12 | 1.23e-06 |

## 4.2 Implementation routes and the degree law

Kernel normalization moves all $1/\gamma$ operator-norm enhancement into classical rescaling: the even kernel $g_2(x) = c\tilde{\gamma}^2/(x^2+\tilde{\gamma}^2)$ realizes $c\gamma^2(A^2+\gamma^2)^{-1}$ with sup $\leq c$ (and the odd kernel $c\tilde{\gamma}x/(x^2+\tilde{\gamma}^2)$ realizes $c\gamma A(A^2+\gamma^2)^{-1}$, sup $\leq c/2$), where $A = H_0 - \omega$, $\tilde{\gamma} = \gamma/(\alpha+\omega)$. Three implementation routes were costed and one is recommended:

- **Route (i), Hermitian decomposition (recommended).** Re$R$ and Im$R$ via odd/even kernels, both analytic (Bernstein poles at $\pm i\tilde{\gamma}$), sup $\leq 1$; per-frequency degree $d = \operatorname{asinh}(\tilde{\gamma})^{-1}\ln(C/\varepsilon')$; $\zeta = \beta^2$.
- **Route (ii), augmented matrix.** The pole-kernel family on the Hermitian augmentation $M = X \otimes A + Y \otimes \gamma I$ realizes a *damped* conjugate resolvent $c\tilde{\delta}(A - i\gamma)(A^2 + \gamma^2 + \tilde{\delta}^2)^{-1}$ (relative error $\tilde{\delta}^2/(\gamma^2 + \tilde{\delta}^2)$ at resonance) — exactness to relative $\varepsilon$ costs an $\varepsilon^{-1/2}$ degree penalty (measured degree ratio 5.98 vs theory 6.00). The corrected version (GSLW Thm 41 standard inversion polynomial) has degree $(s_M/\kappa_M)\ln(1/\varepsilon)$ with $\kappa_M = \sqrt{\text{dist}(\omega,\text{spec})^2 + \gamma^2}$: resonance-matched to route (i), *cheaper off resonance* (adaptive-resolution bonus); retained as the off-resonance alternative.
- **Route (iii), LCU composition.** Dominated (subnormalization pays $(\alpha + \omega + \gamma)$ or $2\times$); not pursued further. Its value is the negative lesson: complex objects must enter through in-kernel poles (i) or Hermitian augmentation (ii), not by applying $(A + i\gamma)$ as an operator.

**Degree law.** The exact geometric coefficient decay of Lemma C.1 (SM §C) gives $d \cdot \operatorname{asinh}(\tilde{\gamma}) = \ln(C/\varepsilon') + O(1)$ with the sufficient constant $C = c/\sqrt{1 + \tilde{\gamma}^2} \leq 0.95$; the archived 30/30 cells (even+odd kernels) define the minimal-degree calibration $C_{\text{emp}} \in [0.882,0.993]$ (a distinct quantity, defined by minimal degrees) — the minimal interpolation degrees track $\ln(c/\varepsilon')/\operatorname{asinh}(\tilde{\gamma})$ within a bounded absolute slack (Fig. 3: $\leq 3$ on the even-kernel cells — diffs $\{1{:}\,8, 2{:}\,3, 3{:}\,4\}$ of 15 — and up to 11 on the coarsest odd-kernel cells, where the minimal search sits well below the truncation bound); high degrees are practical (pyqsp: $d = 1830$ solved in 13.2 s; QSP-vs-polynomial residual $1.4 \times 10^{-4}$ equal to the capitalization prediction to 9 significant digits; Krylov cross-checks at $10^{-15}$; Table 2). The erf window is an entire function and its degree grows only as $\sqrt{\ln(1/\varepsilon)}$ — but its *resolution* semantics differ; the fair comparison is Sec. 4.6.

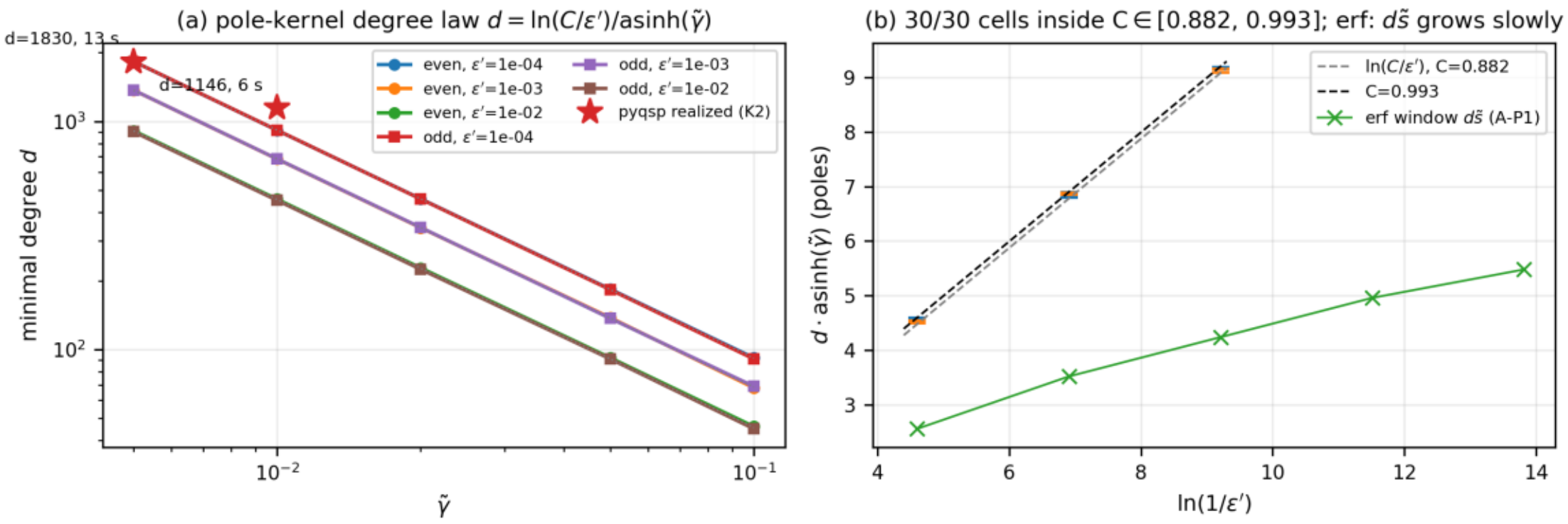

*Figure 3. QSP degree law. (a) Minimal pole-kernel degrees vs $\tilde{\gamma}$ across target accuracies (even/odd kernels), with the law $d = \ln(C/\varepsilon')/\mathrm{asinh}(\tilde{\gamma})$ ($C = 0.94$) and the two realized pyqsp probes (K2; $d = 1830$ in $13.2$ s). (b) Collapse of all 30 cells into the $C \in [0.882, 0.993]$ band; the erf window's $d\tilde{s}$ grows only as $\sqrt{\ln(1/\varepsilon)}$ (different resolution semantics; compared fairly in Fig. 5).*

**Table 2. pyqsp toolchain checks** (C2 = QSP realization vs target polynomial, dominated by the default capitalization; C4/C5 = independent Krylov and matrix-level cross-checks).

| target | degree | phase solve (s) | C2 (QSP vs poly) | C4 (Krylov) | C5 (matrix) |
|---|---|---|---|---|---|
| even Lorentz kernel ($\tilde{\gamma}$=0.02, peak 0.95) | 200 | 0.22 | 9.86e-05 | 1.06e-15 | 5.25e-14 |
| erf window (half 0.5, s=0.02, top 0.95) | 200 | 0.21 | 1.45e-04 | 2.66e-15 | 2.03e-14 |
| odd smooth-inversion kernel ($\tilde{\delta}$=0.02, top 0.95) | 199 | 0.22 | 6.82e-05 | 6.94e-16 | 2.64e-14 |

### 4.3 Theorem 1: SNR-parametrized per-frequency cost, and the error budget

**Theorem 1 ($n = 1$, per frequency).** Assume A1. Let $g_2$ be the even kernel ($c = 0.95$), $U_{g_2}$ its QSP block encoding via Lemma Q, and $\tilde{\mu}_2(\omega) \equiv \langle\psi_0|D\, g_2(A/(\alpha+\omega))\, D|\psi_0\rangle = c\,\bar{S}_\gamma(\omega) \in (0, \beta^2]$ (case $D' = D$; $\bar{S}_\gamma$ is the dictionary object of Sec. 4.1 — the physical spectrum is $S_\gamma^{\mathrm{phys}} = \bar{S}_\gamma/\gamma$). For any $\varepsilon_{\mathrm{rel}} \in (0,1)$ there is an algorithm outputting $\hat{\bar{S}}_\gamma$ with $|\hat{\bar{S}}_\gamma/\bar{S}_\gamma - 1| \le \varepsilon_{\mathrm{rel}}$ using

$$U_H \text{ queries: } \tilde{O}\left(\frac{\alpha+\omega}{\gamma}\cdot\frac{\beta^2}{\tilde{\mu}_2(\omega)}\cdot\frac{1}{\varepsilon_{\mathrm{rel}}}\right), \qquad \text{dipole queries: } O\left(\frac{\beta^2}{\tilde{\mu}_2(\omega)}\cdot\frac{1}{\varepsilon_{\mathrm{rel}}}\right).$$

(The $\tilde{O}$ absorbs $\ln(C/\varepsilon')$ with the instance-independent sufficient constant $C = c/\sqrt{1+\tilde{\gamma}^2} \le 0.95$ of Lemma C.1 — distinct from, and validated by, the empirical minimal-degree calibration constant $C_{\mathrm{emp}} \in [0.882, 0.993]$ of the archived 30 cells (Fig. 3; $C_{\mathrm{emp}}$ is defined by the minimal degrees, not by the sufficient bound, so it may exceed $C$) — and AE logarithms; proof in SM §A. The theorem estimates $\bar{S}_\gamma = \tilde{\mu}_2/c$; for the physical $S_\gamma^{\mathrm{phys}}$ at additive $\varepsilon$, substitute $\varepsilon_{\mathrm{rel}} = \varepsilon\gamma/s^*$ at the peak — the Lorentzian-object corollary of Sec. 6 uses exactly this conversion.) Any positive signal probability is admissible under this SNR parametrization; at resonance $\tilde{\mu}_2(\omega^*) \approx c\,s^*$ (measured $\tilde{\mu}_2/s^* = 0.9538$ $H_2O$-z, 0.9559 LiH-z), so the theorem reads $\tilde{O}((\alpha/\gamma)(\beta^2/s^*)\varepsilon_{\text{rel}}^{-1})$ — structurally identical to the tight-caliber reading of Kharazi et al.'s Theorem 2 (Sec. 4.4), differing only in the $\zeta$ caliber (next paragraph).

**Corollary (amplitude estimation with the square-root kernel).** *The even kernel is nonnegative, so $D\, g_2(A/(\alpha+\omega))\, D$ is positive semidefinite ($D' = D$), and $(\sqrt{g_2})^2 = g_2$. Let $U_{\sqrt{g_2}}$ QSP-realize $\sqrt{g_2} = \sqrt{c}\,\tilde{\gamma}/\sqrt{x^2+\tilde{\gamma}^2}$ (branch points $\pm i\tilde{\gamma}$ as for $g_2$;* ***Lemma C.2 gives the explicit tail bound $|b_k| \le A\sqrt{k}\,\rho^{-k}$ for $k \ge k_0$ with an instance-independent constant****, so the same asinh law holds up to a displayed $\sqrt{d}$ logarithmic correction — empirically the minimal degrees land at $C_{\sqrt{g_2}} \in [0.207, 0.352]$ (SM §H, in-script table); headroom $\sup\sqrt{g_2} = \sqrt{c} < 1$), and $W = U_{\sqrt{g_2}} U_D$ with $\Pi_0$ the all-ancilla projector. The postselection probability on input $|0\rangle|\psi_0\rangle$ is exactly $p = \|\Pi_0 W|0\rangle|\psi_0\rangle\|^2 = \tilde{\mu}_2/\beta^2$, and amplitude estimation in the angle caliber ($M$ controlled calls resolve the Grover-rotation angle to $O(1/M)$ — the phase-readout form of [21] Thm 12 — giving $|\hat{p} - p| \le 2\sqrt{p(1-p)}\;O(1/M)$) estimates $\tilde{\mu}_2$ to relative $\varepsilon_{\text{rel}}$ with $M = \Theta(\beta/(\varepsilon_{\text{rel}}\sqrt{\tilde{\mu}_2}))$: total $U_H$ cost $\tilde{O}(((\alpha+\omega)/\gamma)\cdot(\beta/\sqrt{\tilde{\mu}_2(\omega)})\cdot\varepsilon_{\text{rel}}^{-1})$ and dipole queries of the same SNR form — an improvement factor $\beta/\sqrt{\tilde{\mu}_2}$ over Theorem 1 (6.95 × $H_2O$-z, 10.61 × LiH-z at resonance; Table 4; kernel allocation $\varepsilon'_{\sqrt{g_2}} = \varepsilon_{\text{rel}}\sqrt{\tilde{\mu}_2}/(6\xi)$; proof in SM §A).*

**Remark (scope of the corollary).** The improvement requires a positive-semidefinite sandwich: the construction transfers to any route that admits an explicitly nonnegative (square-form) filter, whereas the standard indicator approximant is not automatically nonnegative, so that use would need a separate construction. Written with the preparation-caliber saving as $\beta/\xi$ it also assumes $\tilde{\mu}_2 = \Theta(\xi^2)$; the general form keeps $\beta/\sqrt{\tilde{\mu}_2}$. For $n \ge 2$ nested objects the sandwich is not positive, so the corollary is confined to $n = 1$.

**Error budget.** The target is the physical response $\text{Im}\alpha^{(1)}(\omega)$, and the budget has four mutually disjoint pieces. (1) AE sampling, from Theorem 1 or the square-root-kernel corollary. (2) The kernel approximation, of size $\varepsilon'\xi^2$: the same scale as the signal rather than $\varepsilon'\beta^2$, a Cauchy–Schwarz tightening, with allocation $\varepsilon' = \varepsilon_{\text{rel}}\tilde{\mu}_2/3\xi^2$. (3) The reversal term $L(-\omega)$ over the $n \ge 1$ lines, handled by the two-frequency protocol $(\omega, -\omega)$ at twice the cost, which also cancels piece 4, whenever the reversal exceeds a budget share $\varepsilon_{\text{sh}}$ of the signal. Corollary 2 gives the condition under the dominant-line approximation, with the next-order $O(\gamma^4)$ correction controlled (SM §A): reversal ≤ share $\varepsilon_{\text{sh}}$ of the signal iff $\gamma \le 2\omega^*\sqrt{\varepsilon_{\text{sh}}/C_{\text{sys}}}$. The 100%-share version, reversal ≤ signal, is the case $\varepsilon_{\text{sh}} = 1$, that is $\gamma \le 2\omega^*/\sqrt{C_{\text{sys}}}$. (4) The $n = 0$ pseudo-line, removed at zero $U_H$ cost by measuring $\langle D\rangle_0$ (zero-cost-certificate lemma below) and subtracting analytically. Pieces (3) and (4) are disjoint because $C_{\text{sys}}$ sums over $n \ge 1$: the $n = 0$ contribution appears once, in piece (4). $C_{\text{sys}}$ itself comes from a few more quantum estimates plus classical post-processing (one peak height, two background points). Measured values and thresholds are in Table 3. At the archived $\gamma$ both systems sit far inside the single-frequency regime (reversal share 0.003%/0.009% over $n \ge 1$), and the two-frequency thresholds lie at or beyond the dominant-line boundary $\gamma \ll \omega^*$ (Table 3, last column). They are extrapolation indicators and are never reached at physical $\gamma$ for these systems.

**Remark (unknown $\tilde{\mu}_2$; off-resonance cost).** If $\tilde{\mu}_2(\omega)$ is unknown, an exponential-doubling schedule ($M = 2^k$, then binary refinement) locates the scale and achieves relative $\varepsilon_{\text{rel}}$ at $\tilde{O}$ overhead (a factor ≤ 2 plus $O(\log(\beta^2/\tilde{\mu}_2))$). Off resonance $\tilde{\mu}_2$ follows the Lorentzian tail, $\tilde{\mu}_2(\omega) \approx c\,s^*/(1 + ((\omega-\omega^*)/\gamma)^2)$, so on a uniform grid of $N = W/\gamma$ points the *average* per-point cost is $\tilde{O}((\alpha/\gamma)(\beta^2/(cs^*))\,[1 + 1/12\,(W/\gamma)^2]\,\varepsilon_{\mathrm{rel}}^{-1})$ — dominated by far-from-peak points under a uniform relative-precision mandate; peak-region-only or additive-precision mandates reduce it back to $\Theta((\alpha/\gamma)(\beta^2/(cs^*))\varepsilon^{-1})$.

**Lemma (zero-$U_H$-cost certificates; A1-caliber).** With the block encoding $U_D$ of $D$ available and the ground state $|\psi_0\rangle$ prepared as a free resource (assumption A1 — in any accounting where each preparation costs $U_H$ queries, these certificates inherit that cost): (i) $\langle D\rangle_0 = \langle\psi_0|D|\psi_0\rangle$ is read off by a Hadamard test of controlled-$U_D$ on $|\psi_0\rangle$— the coin probability is $P(0) = 1/2\,(1 + \langle D\rangle_0/\beta)$, since the top-left block of $U_D$ is $D/\beta$; (ii) $\xi^2 = \|D|\psi_0\rangle\|^2 = \beta^2 P_{\mathrm{ps}}$, where $P_{\mathrm{ps}}$ is the probability of post-selecting the block-encoding ancilla on $|0\rangle$ after applying $U_D$ to $|0\rangle|\psi_0\rangle$. Both use $U_D$ calls only — **zero $U_H$ queries**; precision $\delta_D$ costs $O(\beta\,\delta_D^{-1})$ copies with AE (Lemma 13's optimal readout), or $O(\beta^2\delta_D^{-2})$ classically. In particular the prepared state $|\hat{\chi}\rangle = D|\psi_0\rangle/\xi$ of Corollary 3 and its certificate $\xi^2$ are both accessible before any Hamiltonian query is spent.

**Table 3. Reversal-term protocol** ($C_{\mathrm{sys}}$ over $n \geq 1$; "rev.@$\gamma$" = exact $n \geq 1$ reversal share of $L(\omega^*)$ at the archived $\gamma$; threshold $\gamma_{\mathrm{sw}} = 2\omega^*/\sqrt{C_{\mathrm{sys}}}$ at the 100%-share level — budget-share thresholds scale as $2\omega^*\sqrt{\varepsilon_{\mathrm{sh}}/C_{\mathrm{sys}}}$. For provenance: the excluded $n = 0$ terms $4s_0/s^*$ are 0.858 ($H_2O$-z) and 81.511 (LiH-z), and the archived all-$n$ $C_{\mathrm{sys}}$ values are 2.91 / 82.67.)

| system | $\omega^*$ | $s^*$ | $\gamma$ | $C_{\mathrm{sys}}$ ($n \geq 1$) | rev.@$\gamma$ ($n \geq 1$) | $\gamma_{\mathrm{sw}}$ ($n \geq 1$) | $\gamma_{\mathrm{sw}}/\omega^*$ |
|---|---|---|---|---|---|---|---|
| $H_2O$-z | 1.0728 | 0.5411 | 0.0079539 | 2.0523 | 0.0028% | 1.4977 Ha | 1.40 |
| LiH-z | 0.1331 | 0.5403 | 0.0023188 | 1.1581 | 0.0087% | 0.2474 Ha | 1.86 |

Both thresholds lie at or beyond the dominant-line validity boundary $\gamma \ll \omega^*$ — the two-frequency protocol is never triggered at physical $\gamma$ for these systems; the thresholds are recorded as extrapolation indicators only.

**Corollary 3 (preparation caliber).** With $D' = D$, use a preparation *circuit* for $|\hat{\chi}\rangle = D|\psi_0\rangle/\xi$ (amplitude-amplified, $O(\beta/\xi)$ $U_D$ per copy; $\xi^2 = \langle d^2\rangle_0$ measurable at zero $U_H$ cost, self-certifying) and Hadamard-test the kernel directly: $P(0) = 1/2\,(1 + \tilde{\mu}_2/\xi^2)$. Cost $\tilde{O}(((\alpha + \omega)/\gamma)\cdot(\xi^2/\tilde{\mu}_2)\varepsilon_{\mathrm{rel}}^{-1})\;U_H$ — a saving $(\beta/\xi)^2$ over Theorem 1 ($\beta/\xi$ under the square-root-kernel readout) — and dipole queries $O(\beta\xi/\tilde{\mu}_2\cdot\varepsilon_{\mathrm{rel}}^{-1})$. The preparation *circuit* is frequency-independent (states cannot be reused across measurements — each AE run re-prepares its own copy); the $n$-order generalization saves $(\beta/\xi)^2$ for all $n$ ($\zeta_B^{(n)} = \xi^2\beta^{n-1}$ vs $\beta^{n+1}$).

## 4.4 Comparison with Kharazi et al.'s Theorem 2: where the extra $\alpha$ lives

Their Theorem 2 as published costs $\tilde{O}(\alpha^2\beta^2/(\gamma\varepsilon))$ per bin height, with two distinct sources of $\alpha$: (a) the indicator degree $\tilde{O}(\alpha/\gamma)$ (real — the window width is physical); (b) the subnormalization $\zeta_\omega = (\alpha + 1 + \omega)\beta^2$ rescaling the AE success amplitude (their Eq. 12 and Theorem 2 proof; SI Eq. 40). By Lemma Q the QSP kernel block encoding carries subnormalization exactly 1, so source (b) is removable: the composite of either route — theirs or ours — is a $(\beta\beta')$-block encoding, and the *tight-caliber reading of their own theorem* costs $\tilde{O}((\alpha/\gamma)\cdot(\beta^2/\mathrm{signal})\cdot\varepsilon^{-1})$, the same form as our Theorem 1. The comparison is therefore a **definitional caliber ratio**: under unified prefixes and a common signal anchor the two calibers differ by exactly $\zeta_\omega/\zeta = \alpha + 1 + \omega$ (49.69 × $H_2O$-z, 7.75 × LiH-z — Table 4), a ratio dominated by $\alpha$ (the archived Table 4 columns use each side's native prefix and anchor, giving 46.35 ×/7.26 ×, which decomposes as $(\alpha + 1 + \omega)\cdot\frac{\alpha}{\alpha+\omega}\cdot\frac{\tilde{\mu}_2}{s^*}$; Fig. 4). The correction applies to *both* routes — it is a fix to the published accounting of indicator binning, not an advantage of the resolvent route; the resolvent route's distinct value is the error structure (exact lineshape, disjoint budget) and the preparation saving (Corollary 3). Under the square-root-kernel readout both sides improve again by their own $\sqrt{\zeta/\text{signal}}$ factor (Sec. 4.3 corollary; Table 4).

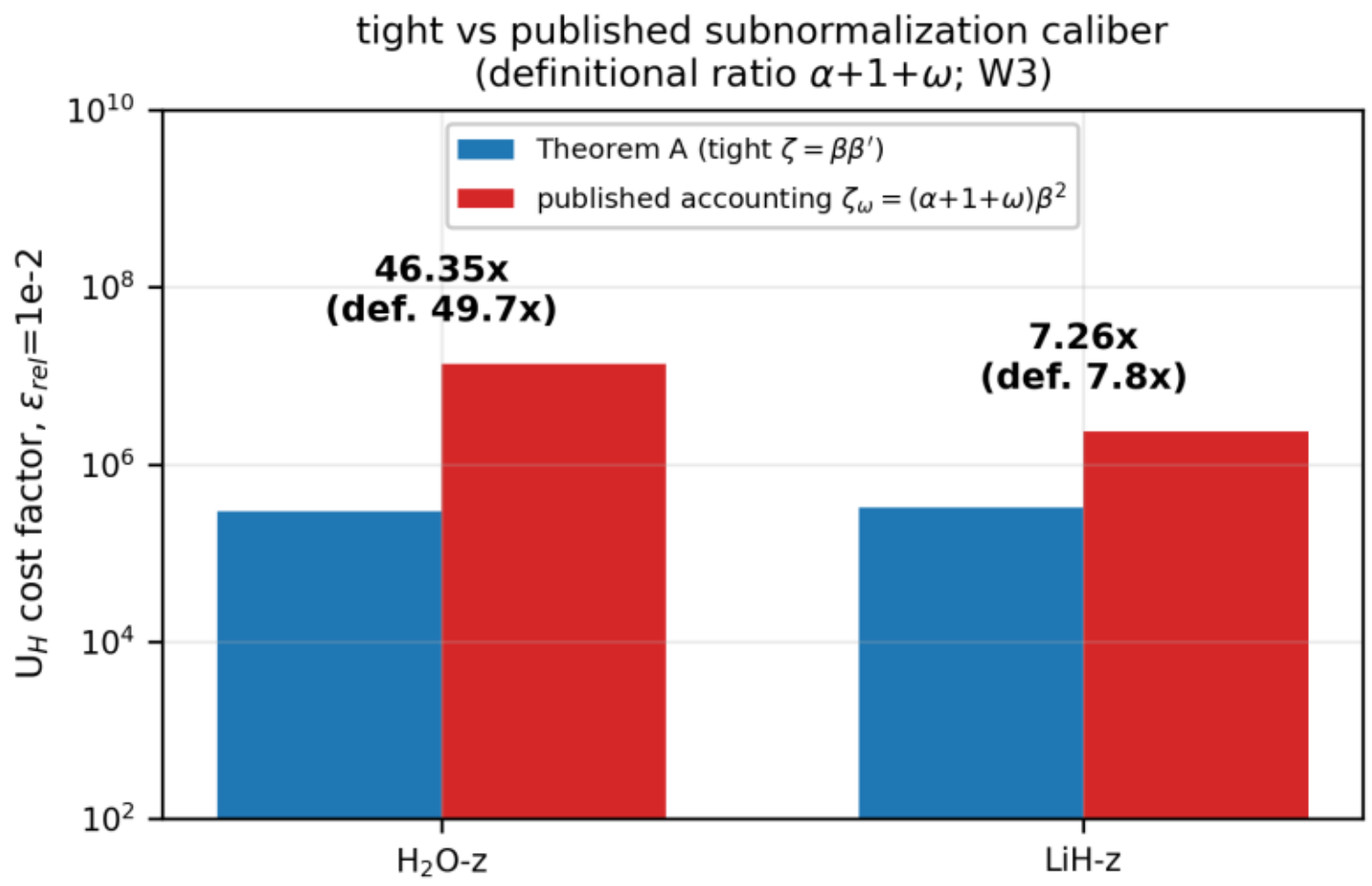

*Figure 4. Tight vs published subnormalization caliber: cost factors at $\varepsilon_{\rm rel} = 10^{-2}$ for $H_2O$-z and LiH-z; labels give the archived mixed-prefix ratio and, in parentheses, the definitional ratio $\alpha + 1 + \omega$ (Sec. 4.4).*

**Table 4. Caliber-ratio certificate** (per-point cost factors as archived — the common factor $\varepsilon_{\rm rel}^{-1}$ (100 at $\varepsilon_{\rm rel} = 10^{-2}$) is omitted from the three cost-factor columns, so ratios are unaffected; "ours"/"paper" keep each side's native prefix and signal anchor as archived; "def." is the unified-prefix, unified-anchor ratio $\alpha + 1 + \omega$; the square-root-kernel variant applies the Sec. 4.3 corollary. Ancillary archived values: $\alpha_{\rm ED} = 47.61/6.62$, $\beta_{\rm ED} = 4.99/7.63$, $\xi^2 = 1.123/12.321$, $s^*/\xi^2 = 48.2\%/4.4\%$.)

| system | $\mu_2/s^*$ | ours, tight | ours, $\sqrt{g_2}$-AE | paper (publ.) | ratio, arch. | ratio, def. |
|---|---|---|---|---|---|---|
| $H_2O$-z | 0.9538 | 2.953e5 | 4.252e4 | 1.369e7 | 46.35 | 49.69 |
| LiH-z | 0.9559 | 3.279e5 | 3.090e4 | 2.382e6 | 7.26 | 7.75 |

### 4.5 Theorem 4 and the $n = 2$ assembly

**Theorem 4 (nested).** With independent ancilla registers and both kernels functions of $A$, the circuit $U = U_D\, U_{g,\omega_2}\, U_D\, U_{g,\omega_1}\, U_D$ (three dipole slots, possibly different directions $d_i$) estimates $\tilde{\mu}^{(2)} = \langle\psi_0|D\, g_2^{(\omega_2)}\, D\, g_2^{(\omega_1)}\, D|\psi_0\rangle$ with $U_H$ cost $\tilde{O}((d_1 + d_2)\cdot(\zeta^{(2)}/|\tilde{\mu}^{(2)}|)\cdot\varepsilon_{\rm rel}^{-1})$, $\zeta^{(2)} = \beta^3$ (preparation caliber $\xi^2\beta$), $d_i = ((\alpha + \omega_i)/\gamma)\ln(C/\varepsilon')$; the SNR parametrization reads $|\tilde{\mu}^{(2)}| = \kappa_2 \cdot$ (strength-chain scale $c^2\Pi_2 S_2$), with the strength chain $S_2$, path count $\Pi_2$, and alignment factor $\kappa_2 \in [0,1]$ given self-contained definitions in SM §B.1. $\kappa_2 = 0$ holds exactly when the object is identically zero — the symmetric zero (Sec. 4.1): centrosymmetric systems with parity-forbidden slot combinations output zero (the algorithm reports $|\hat{\mu}| \le$ noise floor; no false positives). For $n \ge 3$ the explicit induction of SM §B.3 gives $\zeta^{(n)} = \beta^{n+1}$(preparation $\xi^2\beta^{n-1}$), per-shot cost $\sum_i d_i$, per-kernel error budget $\varepsilon_i' = \varepsilon_{\rm rel}|\tilde{\mu}^{(n)}|/(3n\, c^{n-1}\beta^{n-1}\xi_1\xi_{n+1})$, and SNR band $\kappa_n$. Full proof (definitions, nested error propagation, confidence) in SM §B.

**Assembly lemma (Lemma 5).** In the Orr–Ward D1 convention (uniform $\gamma$; $T = 0$) each permutation term of the SOS formula for the first hyperpolarizability equals one nested kernel expectation, $T(i;j,k) = \sum_{m,n} \frac{d^i_{0m} d^j_{mn} d^k_{n0}}{(\omega_{m0}-\omega_\sigma-i\gamma)(\omega_{n0}-\omega_2-i\gamma)} = \langle\psi_0|D_i f_{\omega_\sigma}(A) D_j f_{\omega_2}(A) D_k|\psi_0\rangle$, and $\beta_{ijk} = K$ [six permutations] + c.c. ($K$ a convention constant, complexity-insensitive). Six bookkeeping items (slot–frequency correspondence; six permutations; c.c. mirror; static channel with the D1 $1/(i\gamma)^2$ singularity; symmetric zeros; damping convention) separate the *object-level* identity from the full physical response — the $D = c\mathbb{1}$ counterexample (all commutator responses vanish while bare resolvent chains do not) shows the distinction is real. Three machine-precision checks (Table 5): C2v selection-rule suppression $7.65 \times 10^{-13}$; degenerate Kleinman consistency $1.71 \times 10^{-16}$; loop-formula vs dense-matrix-product anchoring $1.32 \times 10^{-13}$; the static-channel indicator $|d_{00}|^3/\gamma_k^2 = 3.95 \times 10^6$ ($H_2O$-z; $\gamma_k = 10^{-4}$ is the W4 suite's kernel width — at the physical archived $\gamma$ the indicator is 625) quantifies the D1 singularity, whose regularized study (D2/D3 conventions) is deferred (Sec. 9 and SM §B). **Table 5.** $n = 2$ **assembly checks** ($H_2O$, $C_{2v}$, degenerate frequency slots $\omega_\sigma = \omega_1 = \omega_2 = 0.01$).

| check | value | meaning |
|---|---|---|
| selection rule (forbidden/allowed) | 7.65e-13 | C2v-forbidden component suppressed on the assembly formula |
| degenerate Kleinman asymmetry | 1.71e-16 | slot-sum consistency at equal frequencies (machine precision) |
| object anchoring (loop vs matrix) | 1.32e-13 | nested-resolvent path consistency ($-i\gamma$ convention) |
| allowed component scale | 332.1 | $B$ scale (atomic units, $e^3 a_0^3/\mathrm{Ha}^2$) |
| $d_{00}(z)$ | −0.3407 | matches the archived $\lvert d_{00}\rvert^2 = 0.1161$ digit-for-digit |
| static-channel indicator $\lvert d_{00}\rvert^3/\gamma_k^2$ ($\gamma_k = 10^{-4}$) | 3.95e+06 | D1 $1/(i\gamma)^2$ singularity magnitude |

A fourth check (non-degenerate frequency slots, $\omega_\sigma \neq \omega_1$) closes the anchoring at $2.07 \times 10^{-13}$ (suite ap4, SM §H).

## 4.6 Kernel selection: no strict dominance, three criteria

At FWHM-matched resolution $\tilde{\rho}$ (Table 6 summarizes; Figs. 5–6; all numbers from the archived numerical results):

**Table 6. Kernel-selection trade-off** (degrees at $\tilde{\rho} = 0.02$; leakage at physical FWHM $\rho = 2\gamma$ — tail masses $|K|$-normalized on the truncated window $[-20\rho, 20\rho]$ of the analytic kernels; the infinite-interval Lorentz values are 29.5%/15.6%/7.9%; shape bias = peak-normalized sup deviation from the physical Lorentz line shape on the ED spectra, FWHM-matched, fine grids; ranges span $H_2O$-z and LiH-z).

| dimension | Lorentz (resolvent) | Gauss | erf window (indicator route) |
|---|---|---|---|

| dimension | Lorentz (resolvent) | Gauss | erf window (indicator route) |
|---|---|---|---|
| degree @ $\tilde{\rho} = 0.02$, $\varepsilon = 10^{-6}$ | 1376 | **576** ($\sqrt{\ln}$ law, $L/G = 2.38$–2.39) | 1202 ($\kappa = 1/4$); 2552 ($\kappa = 1/8$) |
| leakage beyond $1\rho/2\rho/4\rho$ | 28.4%/14.2%/6.4% (power-law) | 1.9%/0/0 | **0/0/0** (compact) |
| shape bias vs physical lineshape | **0** (kernel is the object) | 13.9–14.6% | 33.2–33.8% |
| readout semantics | Im$R$ (absorption spectrum) | Gaussian-convolved spectrum | flat-top bin average |
| bisection compatibility | weak (no flat top) | weak | **strong** (flat-top semantics) |

**Three criteria.** (1) *Quantitative lineshape* (physical $\gamma$ target) → Lorentz (unbiased; degree premium 2.38 × over Gauss acceptable). (2) *Peak location / resolution of dense lines* → Gauss (cheapest, smallest leakage, closest shape — independent support for [6]); at equal degree $d^* = 1376$ it reaches $\rho_f = 0.0084$ (2.4 × finer) and resolves 7/7 line pairs vs 6/7, but pays the worst fixed-target misfit (0.663 vs 0.050): resolution must be matched to the task. (3) *Bin intensities + bisection* → erf (the only flat-top semantics). Measured isolated-line $L^2$ at matched $\tilde{\rho}$: Lorentz 0.050–0.051 (its $c$-peak floor; a peak-1-target recomputation gives 0.0059, SM §H), Gauss 0.226–0.241, erf 0.311–0.323; erf window sums are exact (0.0%) while peak-calibrated Lorentz/Gauss strengths sit at 1–3%. Peak-normalized kernels at matched FWHM have different areas ($L \approx 3.1\gamma$, $G \approx 2.1\gamma$, $E = 2.0\gamma$ exactly $2H$): the density-estimation readout must divide by the polynomial's own numerical area (measured $0.031216/0.021289/0.019958$ vs analytic $\pi\rho/2 = 0.031416$, $\rho\sqrt{\pi/4\ln 2}$, $2H$).

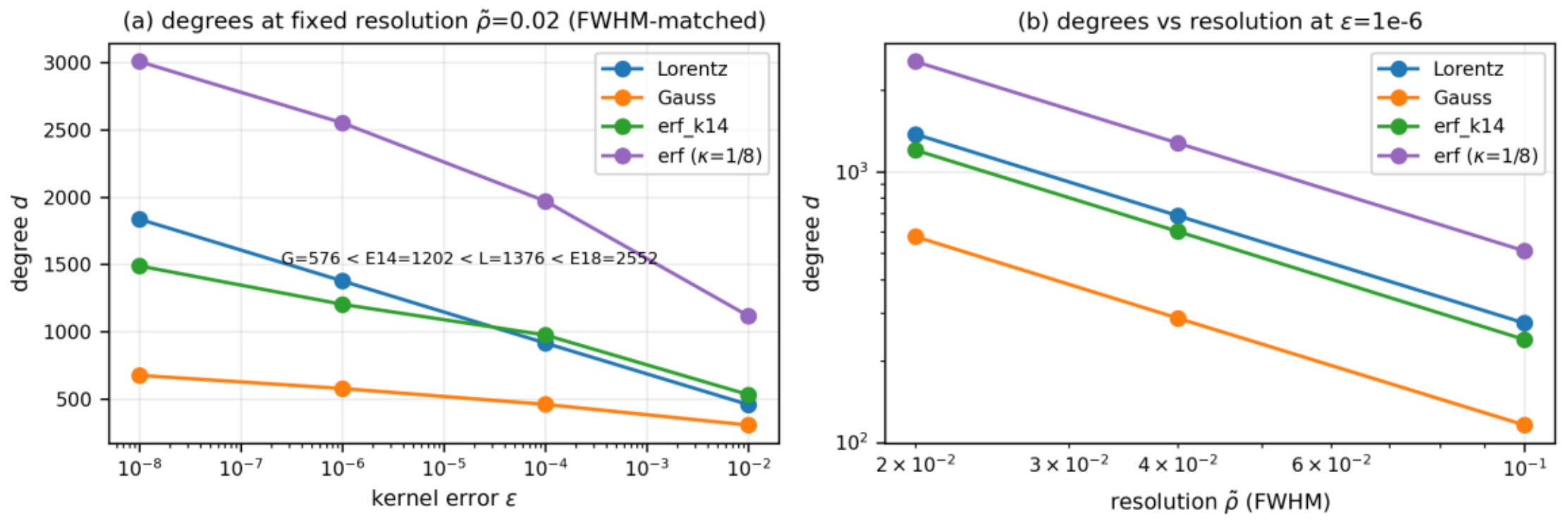

*Figure 5. Kernel-degree Pareto front. (a) Degrees vs kernel error at fixed resolution $\tilde{\rho} = 0.02$ (anchor degrees annotated); (b) degrees vs resolution at $\varepsilon = 10^{-6}$.*

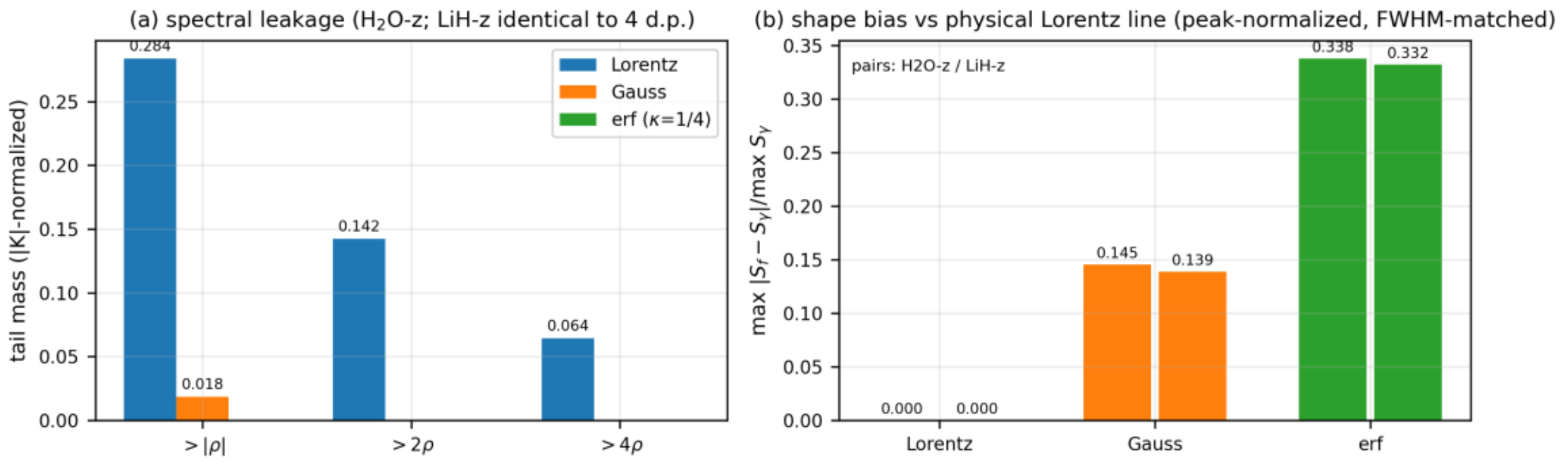

*Figure 6. Leakage and shape bias. (a) Tail mass beyond $1\rho/2\rho/4\rho$ ($H_2O$-z; LiH-z identical to 4 d.p.); (b) shape bias against the physical Lorentz line shape, both systems, peak-normalized FWHM-matched caliber.*

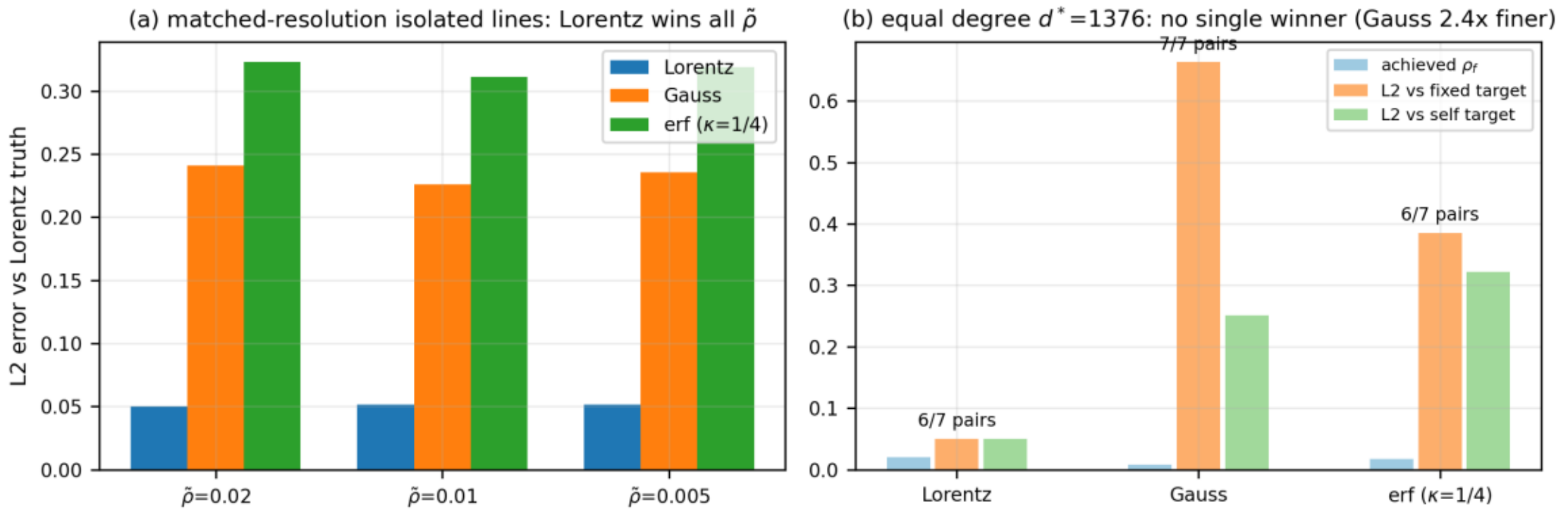

*Figure 7. Isolated-line reconstruction (suite E1; SM §H). (a) Matched-resolution $L^2$ by kernel across three $\tilde{\rho}$; (b) equal-degree ($d^* = 1376$) dual view: achieved resolution, fixed-/self-target fidelities, resolved line pairs (6–7 per the archived gap geometry).*

## 4.7 Continuous spectra and the deconvolution duality

For spectral measures that are densities $\rho(\omega)$, as in lattice dynamical models, the machine behind Theorem 1 produces the measurement-error deconvolution model $y_j = (K_\gamma * \rho)(\omega_j) + \eta_j$, with $K_\gamma$ Lorentzian or Gaussian and $\eta_j$ the AE noise. This is a fixed-design inverse-regression model with observation noise under a known supersmooth operator. The ill-posedness is that of density deconvolution [35], and the errors-in-variables literature [37] is the nearest regression-type analogue, though its covariate-noise design is noisier than ours.

The Lorentzian transform decays as $e^{-\gamma|k|}$, which places it in Fan's supersmooth class. Spectrally truncated deconvolution then has an interior optimal bandwidth $k^*(\varepsilon)$. Sobolev-class densities are recoverable only at logarithmic rates $\|\hat{\rho} - \rho\| \lesssim (\ln 1/\varepsilon)^{-s}$. For densities whose own transform decays exponentially, the analytic class $e^{-a|k|}$, the product transform decays as $e^{-(\gamma+a)|k|}$ and polynomial rates $\sim \varepsilon^{a/(\gamma+a)}$ are attainable. The impossibility of polynomial rates is therefore a statement about the Sobolev class on this data model, not about every prior.

Numerically (Fig. 9, Table 9), $k^*$ grows monotonically and without bound in $\ln(1/\varepsilon)$, and each $\varepsilon$ shows the bias–variance U-curve whose optimum lies inside the pre-asymptotic envelope $1/(\gamma + s/k^*)$; the asymptotic $1/\gamma$ is not reached in the tested range. The static-channel D1 singularity follows $B_{zzz} \sim 6d_{00}^3/\omega^2$ to 0.51%. **Structural correspondence (heuristic).** The quantum degree $d \approx ((\alpha + \omega)/\gamma)\ln(C/\varepsilon')$ and the classical recoverable bandwidth $k^* \sim (1/\gamma)\ln(1/\varepsilon)$ (asymptotic; pre-asymptotic $1/(\gamma + s/k^*)$) both descend from the same exponentially decaying kernel transform — a structural correspondence *suggestive of a common information budget*. We do not claim a theorem transfer: Fan's rates hold under their own model assumptions (sampling design, bounded independent noise, density smoothness class, boundary conditions), our data model differs (the quantum polynomial-approximation error and the classical measurement/deconvolution noise are distinct error parameters), and similar $(1/\gamma)\log(1/\varepsilon)$ factors in two formulas do not by themselves establish an information-saturation statement. The asymptotic $1/\gamma$ regime is not reached in the tested $\varepsilon$ range (below). Kernel-vs-bin for continuous spectra (Fig. 8, Tables 7–8): under the **per-measurement-degree-matched caliber** (equal degree *per measurement* — a per-circuit depth caliber) the kernel arm is more accurate ($L^2$ 0.0149 vs 0.0294) *while using* $2.78 \times$ *more total degree* (28800 vs 10368) and more AE repetitions ($5.0 \times 10^4$ vs $1.8 \times 10^4$) — this cell carries **no** total-resource claim. The total-resource conclusion rests on the **equal-resolution row alone**: $2.8 \times$ fewer total degrees (28800 vs 81178) at equal AE repetitions ($5.0 \times 10^4$) *and* $7 \times$ more accurate ($L^2$ 0.0149 vs 0.108). Bin noise grows with bin count (0.031 → 0.424 for $m = 18 \to 200$) while the kernel arm improves (0.021 → 0.0096; Table 8) — the degree-budget analogue of per-bin signal-to-noise threshold behavior.

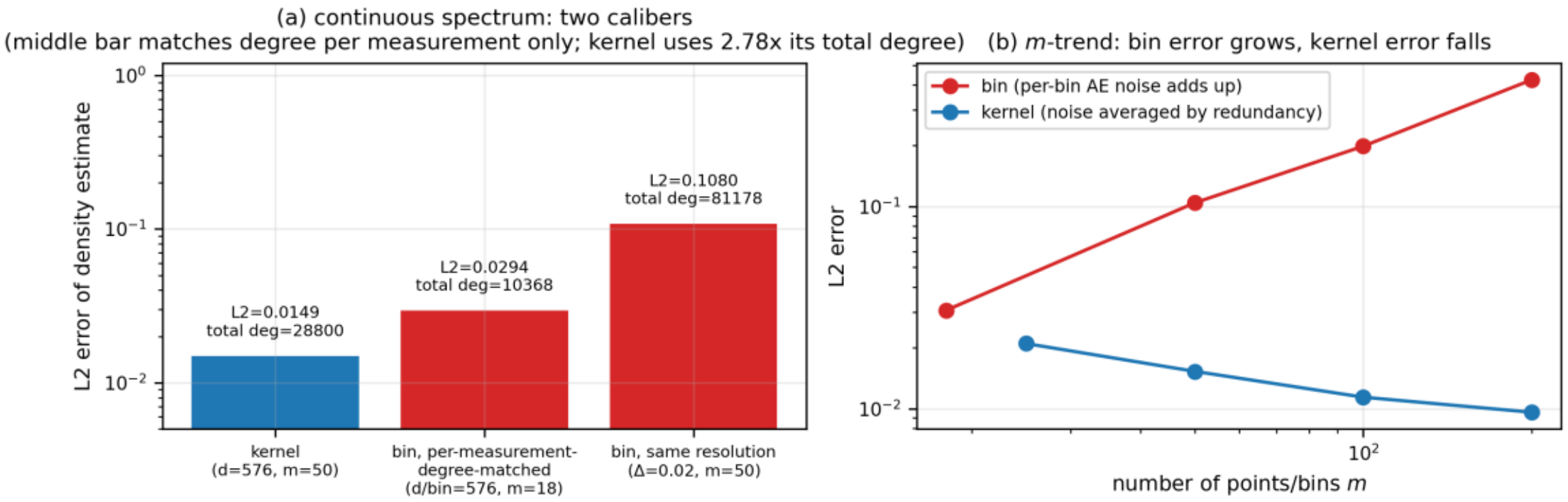

*Figure 8. Continuous-spectrum comparison (suite E2; SM §H). (a) Kernel vs bin under the per-measurement-degree-matched caliber (circuit-depth caliber: equal degree per measurement; the kernel entry uses* $2.78\times$ *more total degree — no total-resource claim attaches to this cell) and under the equal-resolution caliber (total degrees annotated; this row carries the total-resource conclusion). (b) The* $m$*-trend: bin error grows (per-bin AE noise adds), kernel error falls (redundant-frame averaging).*

**Table 7. Continuous-spectrum comparison with resource columns** (degrees per measurement and totals as archived independently; per-point noise $\varepsilon_{\mathrm{rel}} = 10^{-3}$; the AE-repetition and dipole-query columns are computed by the stated readout caliber — AE repetitions $m \cdot \varepsilon_{\mathrm{rel}}^{-1}$ per Lemma 13 with $\Theta(1)$ signals, dipole queries 2 per AE repetition from the $D'PD$ sandwich; no new experiments; † matched in degree *per measurement* only — a circuit-depth caliber; this cell carries no total-resource claim).

| scheme | m | degree/meas. | total degree | AE rep. | dipole q. | $L^2$ |
|---|---|---|---|---|---|---|
| kernel Gauss d=576 | 50 | 576 | 28800 | 5.0e+4 | 1.0e+5 | 0.0149 |
| bin, per-meas. degree-matched † | 18 | 576 | 10368 | 1.8e+4 | 3.6e+4 | 0.0294 |
| bin, same-resolution | 50 | 1624 | 81178 | 5.0e+4 | 1.0e+5 | 0.1080 |

**Table 8.** $m$**-trends** (bin degrees adaptive as archived).

| m | degree/bin | total degree | bin $L^2$ | kernel $L^2$ (d=576) |
|---|---|---|---|---|
| 18 | 584.5 | 10521 | 0.0307 | — |
| 25 | — | — | — | 0.0210 |
| 50 | 1623.6 | 81178 | 0.1043 | 0.0153 |
| 100 | 3247.1 | 324710 | 0.1989 | 0.0114 |
| 200 | 6494.2 | 1298841 | 0.4237 | 0.0096 |

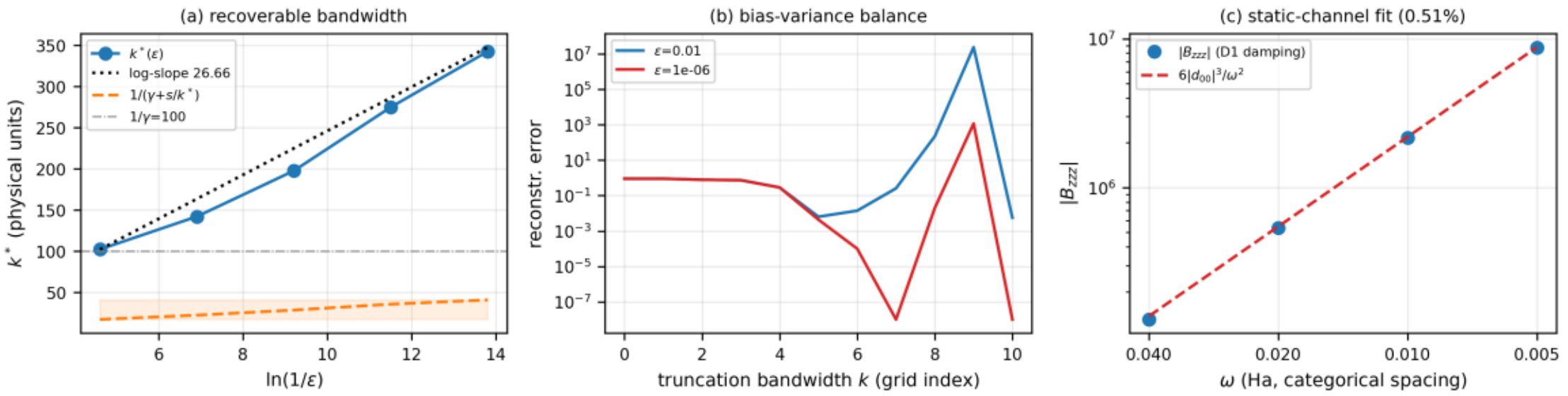

*Figure 9. Supersmooth deconvolution (suite D1/D2; SM §H). (a) Optimal bandwidth $k^*(\varepsilon)$ with fitted log-slope and the pre-asymptotic envelope (recomputed in-script from $1/(\gamma+s/k^*)$, $\gamma=0.01$, $s=5$); (b) bias–variance U-curves at two noise levels; (c) the static-channel singularity $B_{zzz}$ vs the $6|d_{00}|^3/\omega^2$ prediction (fit 0.51%).*

**Table 9. Deconvolution bandwidth** (the archived values are rounded to four decimals, so the two smallest entries are below the archived precision floor and are reported as below $5\times10^{-5}$ rather than as zeros).

| $\varepsilon$ | $k^*$ | min err |
|---|---|---|
| 1e-02 | 102.0 | 5.7e-03 |
| 1e-03 | 141.9 | 9.0e-04 |
| 1e-04 | 197.5 | 2.0e-04 |
| 1e-05 | 274.8 | <5e-05 |
| 1e-06 | 342.5 | <5e-05 |
| log-slope 26.66 | envelope [16.9, 40.7]; asymptote $1/\gamma=100$ | |

## 5 The time-domain route

### 5.1 The single-point AE theorem and the grid self-consistency checks

**Theorem 6 ($n$-time correlation functions).** With $U_H$, $U_D$ as above and $t_j\in[0,T]$, estimate $C_n(t_n,\dots,t_1)=\langle\psi_0|D\,e^{-iH_0t_n}D\cdots D\,e^{-iH_0t_1}D|\psi_0\rangle$ by a controlled-$A$ Hadamard test with QSP-simulated evolution segments and AE. **(a) Single point** ($T_{\text{tot}}=\sum_j t_j$): dipole queries $O(n\beta^{n+1}/\varepsilon)$; $U_H$ queries $\tilde{O}(\alpha\beta^{n+1}T_{\text{tot}}/\varepsilon)$; $O(\beta^{n+1}/\varepsilon)$ circuit copies; all at confidence $1-\delta$ via median-of-$R=\lceil 6\ln(2/\delta)\rceil$ repetitions (simulation error $\delta_{\text{sim}}=\varepsilon/4n$ budgeted; SM §D.1). **(b) Grid full-spectrum, worst-case sampled-noise accounting** (Lorentz damping $\gamma\le 0.127\,W$; window $T=\lceil L/\gamma\rceil$ with the window logarithm $L:=\ln(4A_1/(\gamma A_s\varepsilon))$ defined in D.2; step $\Delta t=\pi/W$; $m=\Theta((W/\pi)(L/\gamma))$ per dimension; per-point additive noise budget $\delta_{\text{pt}}=\varepsilon A_s/(2T^n)$; scales $A_1$, $A_s$ defined in SM §D.2): the trapezoid reconstruction has $L^\infty$ accuracy $\varepsilon A_s$ over the window provided $\varepsilon\ge\varepsilon_{\text{floor}}:=4.8A_1/(WA_s)$ — the worst-case aliasing floor of SM §D.2 (band-edge geometry; for $\varepsilon<\varepsilon_{\text{floor}}$ refine $\Delta t$ by the factor $r=\Theta(\varepsilon_{\text{floor}}/\varepsilon)$, which grows the per-dimension count $m\to rm$ and multiplies the total grid — hence the total cost — by $\Theta(r^n)$; the one-power form is the $n=1$ special case) — using

$$U_H \text{ queries } n(W/\pi)^n\alpha\beta^{n+1}L^{2n+1}/(\gamma^{2n+1}\varepsilon A_s)\times O(n\ln m+\ln(1/\delta)),$$

which at $W = 2\alpha$ becomes $n2^n\alpha^{n+1}\beta^{n+1}L^{2n+1}/(\pi^n\gamma^{2n+1}\varepsilon A_s)$ (the centered-encoding specialization $W = 2\alpha$; the uncentered encoding of the numerics, $\alpha_{\mathrm{ED}} = W$, replaces $2\alpha$ by $W$). *Instance-family statement (the two readings of $A_s$ made precise; SM §D.2).* The displayed $\gamma^{-(2n+1)}$ and the logarithm $L = \ln(4A_1/(\gamma A_s\varepsilon))$ belong to **family F2** (fixed target scale; variants F2a pure task normalization — $A_s, A_1$ fixed, $L$ tracks $\gamma$ — and F2b rescaled-spectrum realization — $A_1 \sim \gamma$, $L$ $\gamma$-free; SM §D.2). **Family F1** (fixed spectrum: all $s_n$, $\omega_n$ fixed while $\gamma \to 0$; then $A_s = s^*/\gamma$ automatically, and readout is relative to the physical peak height $s^*/\gamma$) gives $1/A_s = \gamma/s^*$, cancelling one $\gamma$-power: total cost $\gamma^{-2n}$ with the **$\gamma$-free** logarithm $L = \ln(4A_1/(s^*\varepsilon))$. The two exponents answer *different asymptotic questions* — they are never mixed within one comparison (the genealogy entries whose $\gamma$-exponent depends on the family are tagged F1/F2 in Sec. 7.2); the "joint $(\gamma, \varepsilon)$ asymptotics must keep $L$" caveat is an F2 statement, vacuous in F1. Relative to a naive per-point allocation $|\eta_\lambda| \le \varepsilon A_s$ without propagation accounting, the deterministic bound of Lemma D.4 raises the cost by $T^n$ ($\gamma^{-(n+1)} \to \gamma^{-(2n+1)}$, $L^{n+1} \to L^{2n+1}$): the per-point budget $\delta_{\mathrm{pt}}$ propagates through the quadrature with amplification $T^n$, which must be bought back through the AE repetitions. *Variant (remark)*: under per-point noise independence (the assumption grade of Proposition 7, not used for the theorem), the amplification is $\lesssim (\Delta t\, T)^{n/2}$ and the $\gamma$-exponent improves to $3n/2 + 1$. Success probability and confidence logarithms are displayed rather than absorbed (per-point median-of-$R$, $R = O(n \ln m + \ln(1/\delta))$, SM §D.3). Proof in SM §D. *Reading.* The per-grid structure carries subnormalization $\zeta_t = \beta^{n+1}$ (no indicators), but coverage plus worst-case noise amplification costs $\gamma^{-(2n+1)}$: relative to the naive accounting the amplification $T^n$ multiplies the grid/bisection bound ratio by $(L/\gamma)^n$, which exceeds 1 for all $n \ge 1$ at both archived parameter sets (a comparison of two upper bounds of different tightness, not of the algorithms themselves). The route's value is the data model and the single-point cost structure, not a competitive grid algorithm.

The grid checks are an independent numerical verification of the analytic bounds of SM §D (Fig. 10, Table 10). They are deterministic, they exercise the implemented quadrature and the error laws, and they do not stand in for the proof. Truncation ratios fall by 0.013–0.029 per half-window step, centred on $e^{-4} = 0.018$. Discretization ratios are 0.245–0.253 per halving, the $(z\Delta t)^2/12$ structure of Lemma D.3(ii). Nyquist violation amplifies the reconstruction error by $1.3 \times 10^3$–$2.2 \times 10^4$. The folded peaks come from the strongest line for $H_2O$ and from a weaker line at 2.47 Ha with $s = 0.0135$ for LiH, so shrinking $W$ below the full support is not free in either case. Resolution valley depths are 3.8–11.7. The grid-point scaling $m = \Theta((\alpha/\gamma) \times$ window scale) is verified mechanically: the G1d check fixes the window at $T = 20/\gamma$ (window constant $c = 20$, a fixed instance of the window scale rather than the theorem's window logarithm $L$ at some $\varepsilon$), for which $m/(\alpha/\gamma) = 2c/\pi = 12.732$ in all three systems. The $n = 2$ reconstruction reproduces the archived nested-resolvent values to $6.9 \times 10^{-6}$–$9.3 \times 10^{-5}$, far below the worst-case aliasing floor of Theorem 6(b) at these parameters, as a worst-case bound should be. The symmetry-zero system rebuilds $|R_2| \sim 10^{-10}$. For the sparsity motivation: the full sto-3g width is $W = 47.61$ Ha ($H_2O$) while the strongest transition sits at $\omega^* = 1.07$ Ha, a 44 × gap (LiH 6.62 vs 0.13, 50 ×). Grids are priced by $W$, spectra live on narrow supports; that is the three-layer motivation (prior window, distribution sampling, sparse recovery) behind Sec. 5.3.

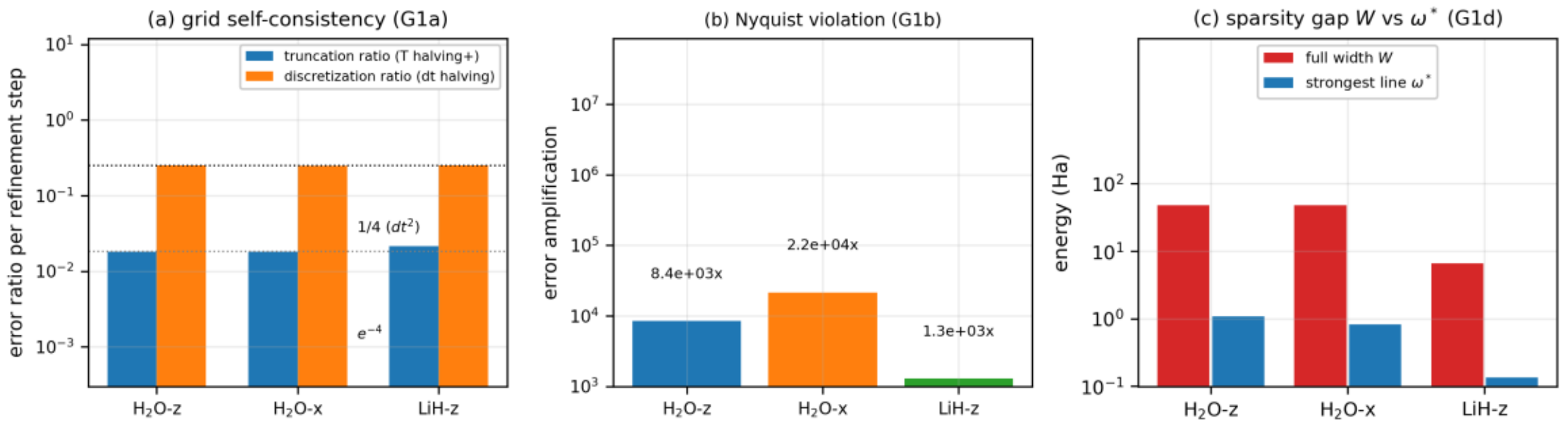

*Figure 10. Time-domain grid self-consistency (suite G1; SM §H). (a) Truncation and discretization error ratios (guides $e^{-4}$ and $1/4$); (b) Nyquist-violation amplification with fold targets; (c) the sparsity gap: full width $W$ vs strongest line $\omega^*$, with the grid points per dimension (fixed window constant $c = 20$: $m/(\alpha/\gamma) = 2c/\pi \approx 12.73$).*

**Table 10. Grid-check summary** (G2 = $n = 2$ time-domain reconstruction vs the archived suite-ap0 V3 nested-resolvent values).

| check | $H_2O$-z | $H_2O$-x | LiH-z |
|---|---|---|---|
| G1a truncation ratio (median) | 0.0182 | 0.0182 | 0.0213 |
| G1a discretization ratio (median) | 0.2520 | 0.2464 | 0.2504 |
| G1b Nyquist amplification | 8.4e+03 | 2.2e+04 | 1.3e+03 |
| G1c resolution valley depth | 11.70 | 5.44 | 3.77 |
| G2 $n = 2$ cross-check | 6.87e-06 (rel) | 4.21e-10 (rel, symmetry zero) | 9.27e-05 (rel) |
| $W_{\text{top}}$ / $\omega^*$ (Ha) | 47.61 / 1.0728 | 47.61 / 0.8259 | 6.62 / 0.1331 |
| $m$ per dim, $m/(\alpha/\gamma) = 12.732$ | 38110 | 38110 | 18172 |

## 5.2 Same-caliber conversion of GQPE (Loaiza et al.)

The conversion chain is recorded in SM §G. We checked it against the journal full text and display the steps here and in the footnotes of Table 14b. Resolution: $\eta_L = \pi\gamma/W$, using their $\|\widehat{H}\| \le \pi$ scaling, with general-$\widehat{H}$ penalties absorbed in $\tau = \pi/W$. Normalization: $\Omega = \prod_j \|V^{(j)}\|_1 \to \beta^{n+1}$. Each scaled $e^{\pm i\widehat{H}}$ step equals one QSP-simulated physical step at $\alpha\tau = \pi/2$ in the centred encoding $\alpha = W/2$, tight at exact spectral fit; with the uncentred encoding used in our numerics, $\alpha_{\text{ED}} = W$, the same step costs $\alpha\tau = \pi$. In either encoding the conservativeness of $\alpha$ is paid, not hidden. The response scaling $R_s = R/\tau$ is easy to miss: a physical $\varepsilon$ task maps to scaled precision $\varepsilon_s = \varepsilon/\tau$.

Three consequences follow. The fixed-point cost at $n$-th order is $(\pi/2)n\beta^{n+1}/(\gamma\varepsilon)\, U_H$ queries, because the $\tau$ factors of $\eta_s$ and $\varepsilon_s$ cancel. No $\alpha$ power is left in this additive-precision caliber on the physical response $R$. The natural scale $|R|_{\text{pk}} = \Theta(\tau s)$ itself carries $\tau = \pi/W$, so in relative-to-peak units $\varepsilon = \varepsilon_{\text{rel}}\tau s$ reintroduces the bandwidth and the cost reads $\Theta(W\beta^{n+1}/(\gamma s\varepsilon_{\text{rel}}))$; the window sits in the time register, outside the block-encoded operator, and no $\alpha$ enters the operator accounting under either semantics. There are also $O(\beta^{n+1}\pi/(2\alpha\varepsilon))$ block-encoding and preparation queries. Full-spectrum output-distribution sampling costs $\gamma^{-1}$ per circuit with $O(\Omega^4/\varepsilon^4)$ samples, or $\Omega^4/\varepsilon^2$ in the complete-square variant. The single-ancilla variant carries the $(C10)$ denominator-side $\tau^{2D-1}$ penalty $(W/\pi)^{2n-2}$.

Two pitfalls are worth recording for later cross-caliber comparisons. The PDF text extraction of Eq. C10 flattens the fraction, so $\tau^{2D-1}$ is easily misplaced; it sits in the denominator, as confirmed from the journal typeset geometry and the arXiv v3 TeX source. And "cost linear in order" holds only for the per-circuit evolution and the register count: $\eta_L^{2D}$ and $\Omega^4$ do not scale linearly.

### 5.3 Conditional sparse-recovery stability, and a bandwidth-decoupled target

**Proposition 7 ($n = 1$, equispaced-window ESPRIT — a classical conditional stability result transferred to the quantum sampling cost model; hypothesis set made explicit).** Assume: (i) spectrum $s$-sparse with nonnegative strengths, support in $[0, W]$, minimum separation $\geq \Delta_{\text{sep}} = 2\gamma$; uniform damping $\gamma$; (ii) per-point estimates $\hat{C}(t_j)$ of the damped correlation $C_2(t) = \sum_k a_k e^{-(\gamma+i\omega_k)t}$ at $m = \Theta(W/\Delta_{\text{sep}})$ **equispaced** times on a window $T = \Theta(1/\Delta_{\text{sep}})$, with additive noise $|\eta_j| \leq \varepsilon_q$ (AE caliber, success $\geq 8/\pi^2$ per point, [21] Thm 12; **independence of the $\eta_j$ across sampling times is an explicit assumption** — structural argument, no counterexample, no proof; Sec. 9; the statement is conditioned on the **joint event** that all $m$ sampling points simultaneously satisfy $|\eta_j| \leq \varepsilon_q$ — per-point median amplification to failure probability $O(\delta/m)$ secures this joint event at overall confidence $1 - \delta$, at a per-point logarithmic cost already absorbed by $\tilde{O}$); (iii) damping conversion and SRF bookkeeping as in SM §F(ii), (iv); (iv) **small-perturbation regime** (the support-recovery criterion of SM §F(i), adopted as a hypothesis): the noise is small enough that the quoted perturbation analysis applies and the induced pole perturbation stays **strictly below** $\Delta_{\text{sep}}/2$ — in symbols, $\varepsilon_q < \varepsilon^*$ with $\varepsilon^*$ an instance threshold parametrizing the hypothesis as a function of the Hankel conditioning (the smallest singular value of the noiseless factorization), the minimum amplitude $a_{\min}$, the separation, and the window length; $\varepsilon^*$ **parametrizes the assumption, it is not derived**; (v) the model order $s$ is known (Hankel rank / order selection). Algorithm: per point run the Theorem 6(a) circuit; then classical damped ESPRIT (Hankel SVD + shift invariance, matrix-pencil family [36, 12]). Then (a) the support is recovered **exactly** — this is SM §F(i)'s criterion, adopted as hypothesis (iv); the proposition packages the quoted stability, it does not derive the criterion; (b) the recovered poles $(\hat{\omega}_k, \hat{\gamma})$ and amplitudes $\hat{a}_k$ obey perturbation bounds **linear in** $\varepsilon_q$, with conditioning factors polynomial in the spectral resolution factor $\text{SRF} = 1/(f_c\Delta_{\text{sep}})$. Total $U_H$ cost $\tilde{O}((W/\Delta_{\text{sep}})\,\alpha\beta^2/(\gamma\varepsilon_q))$ — the sample count here is tied to the bandwidth (this caliber is *not* a reduction over the grid; its value is the validated error law and the off-grid accuracy). The statement is a **conditional stability proposition quoted from the classical literature** plus our damping conversion — not a self-contained recovery theorem; SM §F lists every quoted condition, including the content of hypotheses (iv)–(v).

**Remark 7.1 (random-subset ANM — a conditional target, not a claim).** Sampling a random subset of $m \geq C\max\{\log^2(n/\delta),\ s\log(s/\delta)\log(n/\delta)\}$ points of the equispaced grid and solving atomic-norm minimization (ANM) would decouple the sample count from the bandwidth ($\gamma$ only in logarithms), with total $U_H$ cost $\tilde{O}(s\,\text{polylog}(s, W/\gamma) \cdot \alpha\beta^2/(\gamma\varepsilon_q))$. This rests on Tang et al. Thm 1.1 [29], whose hypotheses include a **random-signs condition that our nonnegative-strength model does not satisfy**; the ANM pipeline is also not verified in our numerics (no SDP solver in the environment — the ESPRIT side carries the numerical evidence). The bandwidth-decoupled sample count is therefore registered as an open verification problem (Sec. 9), and the formula above is quoted *under those unverified conditions*, not proved here.

(The statement is conditional: hypotheses (i)–(v) — in particular the perturbation-regime threshold $\varepsilon^*$, which parametrizes rather than derives the support-recovery criterion — are adopted from the classical analyses, and conclusion (a) restates hypothesis (iv); the contribution here is the damping conversion and the quantum-side cost.)

The numbers are in Fig. 11 and Tables 11–12 (16 cores, seed 20261023). The ESPRIT error is linear in $\varepsilon_q$: tenfold-precision ratios are 9.97–10.07, median 10.00. The plateau err/$\varepsilon_q \approx 0.45$–0.50 holds across $s = 3,5,10$ for $m \in [16s, 64s]$. SRF doubling gives a median growth of 5.91 (the archived `pair_b_median_growth` statistic over ten seed pairs, p90 76), and one catastrophic seed is disclosed: err/$\varepsilon_q = 508$ at $s = 10$. The bounded-noise control at $s = 5$, $m = 160$ has median 0.448 over 20 seeds against 0.338 for the Gaussian model over 10 seeds, so the AE-caliber noise is the stricter one; a same-seed rerun gave 0.473, a run-log value outside the archived results (SM §H). Two-dimensional tensor-product line spectra, built as true outer products, recover their supports in 20/20 seeds, with nearest-line distances $\sim 3 \times 10^{-5} \ll 2\gamma$. The ANM sample-count law ($m \sim s$ polylog) is cited rather than verified numerically, since no SDP solver was available here; the ESPRIT side carries the numerical evidence, and the caliber difference is registered in Sec. 9.

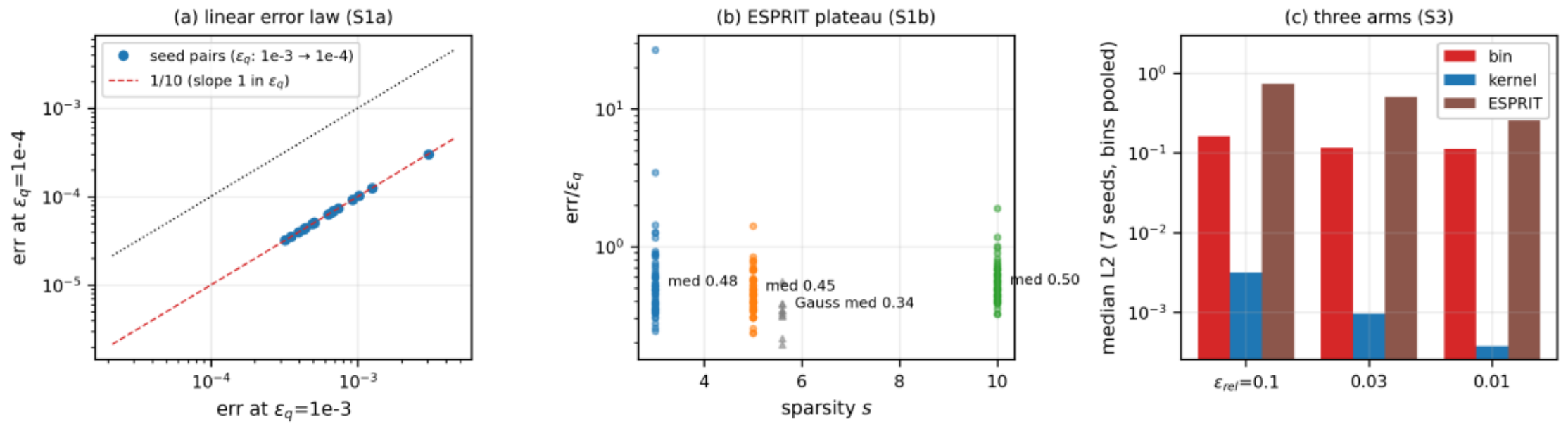

*Figure 11. Sparse recovery (suite S1/S2; SM §H). (a) The linear error law across seed pairs ($\varepsilon_q$: $10^{-3} \to 10^{-4}$; median ratio 10.00); (b) the ESPRIT plateau across sparsities with the Gaussian-vs-bounded noise contrast; (c) the three arms on the 6-line Lorentzianized spectrum (medians over 7 seeds, bins pooled).*

**Table 11. Sparse-recovery statistics** (suites S1/S2; SM §H).

| experiment | statistic | value |
|---|---|---|
| S1a error linearity | ratio range across 20 seed pairs (median) | 9.97–10.07 (10.00) |
| S1b plateau $s = 3$ | median err/$\varepsilon_q$, $16s \le m \le 64s$ | 0.485 |
| S1b plateau $s = 5$ | median err/$\varepsilon_q$, $16s \le m \le 64s$ | 0.450 |
| S1b plateau $s = 10$ | median err/$\varepsilon_q$, $16s \le m \le 64s$ | 0.498 |
| S1b SRF doubling | archived pair_b upper-median statistic (p90) | 5.91 (76) |
| S1c noise model | median err/$\varepsilon_q$ at $s = 5$, $m = 160$: Gaussian (10 seeds) vs bounded (20 seeds) | 0.338 vs 0.448 |

| experiment | statistic | value |
|---|---|---|
| S2 2D support | success 20/20; nearest-line distance median (A/B) | 3.5e-05 / 3.1e-05 |

**Table 12. Three-arm medians on the 6-line Lorentzianized spectrum** (suite S3; 7 seeds per configuration).

| n_bins | $\varepsilon_{\text{rel}}$ | med L2 bin | med L2 kernel | med L2 ESPRIT |
|---|---|---|---|---|
| 16 | 0.1 | 0.2286 | 0.00315 | 0.6017 |
| 16 | 0.03 | 0.2045 | 0.00119 | 0.6053 |
| 16 | 0.01 | 0.1979 | 0.00040 | 0.2756 |
| 32 | 0.1 | 0.1621 | 0.00380 | 0.7922 |
| 32 | 0.03 | 0.1154 | 0.00082 | 0.3972 |
| 32 | 0.01 | 0.1121 | 0.00044 | 0.2540 |
| 64 | 0.1 | 0.1302 | 0.00259 | 0.7337 |
| 64 | 0.03 | 0.0672 | 0.00091 | 0.5773 |
| 64 | 0.01 | 0.0568 | 0.00032 | 0.1994 |

## 6 Query lower bounds

All lower bounds in this section are stated in the resource models of SM §E.1 — total evolution time, circuit depth (serial evolution time), oracle-call count, circuit copies — kept separate throughout; the oracle/probe model (controlled evolution oracle, arbitrary adaptive durations, free probe preparation, ancillas allowed) is defined there.

**Objects and their $\varepsilon$-semantics.** The windowed object $R_{\text{win}}$ (Loaiza Thm 1's windowed series; peaks $O(s) \leq \Omega$) and the Lorentzian object $R_L = S_\gamma^{\text{phys}}$ (ours; peak height $\sim s/\gamma$) carry *different precision semantics*: their instance-pair sensitivities differ by a full factor $1/\gamma$ ($\pi/2\gamma$ vs $18/(16\sqrt{3}) \cdot 1/\gamma^2 = 0.6495/\gamma^2$; both measured to analytic precision, Table 13), and each object supports its own legitimate — and matched — lower bound. Attaching an instance pair to the object whose slope you then use is the only rule; bounds on different objects cannot contradict each other. For the Lorentzian object this yields a new corollary with exponent-matched bounds:

**Corollary 15 (Lorentzian-object exponent matching).** *For instance families with $s^* = \Theta(1)$: estimating $R_L(\omega)$ to additive $\varepsilon$ requires $T_{\text{total}} = \Omega(1/(\varepsilon\gamma^2))$ (Lemma E1 with the max-slope pair $\delta^* = (2+\kappa)\varepsilon\gamma^2/(0.6495\, s^*)$; SM §E.3), and Theorem 1 run at $\varepsilon_{\text{rel}} = \varepsilon\gamma/s^*$ achieves $\tilde{O}(\alpha\beta^2/(\varepsilon\gamma^2))$ (Sec. 4.1 conversion dictionary) —* ***matching the $\gamma$ and $\varepsilon$ exponents across the stated oracle models*** *(the lower bound lives in the total-evolution-time caliber of E.1, the upper bound in $U_H$ query count; a same-model match would additionally require an evolution-oracle realization of the resolvent route — e.g. a phase-estimation-based block encoding of $H_0$ — registered as future work per Lemma E.0(ii)). Under relative-to-peak precision the same object recovers the $\Theta(1/(\gamma\varepsilon_{\text{rel}}))$ form, matching the windowed task's scaling; the two objects differ exactly by the peak-height $1/\gamma$ enhancement, not by any "mismatch" (SM §E.3).*

The $\varepsilon$-semantics dictionary (additive-on-$R_L$ / relative-to-peak / additive-on-$R_{\text{win}}$) is carried into Tables 14a/14b and the genealogy of Sec. 7.2.

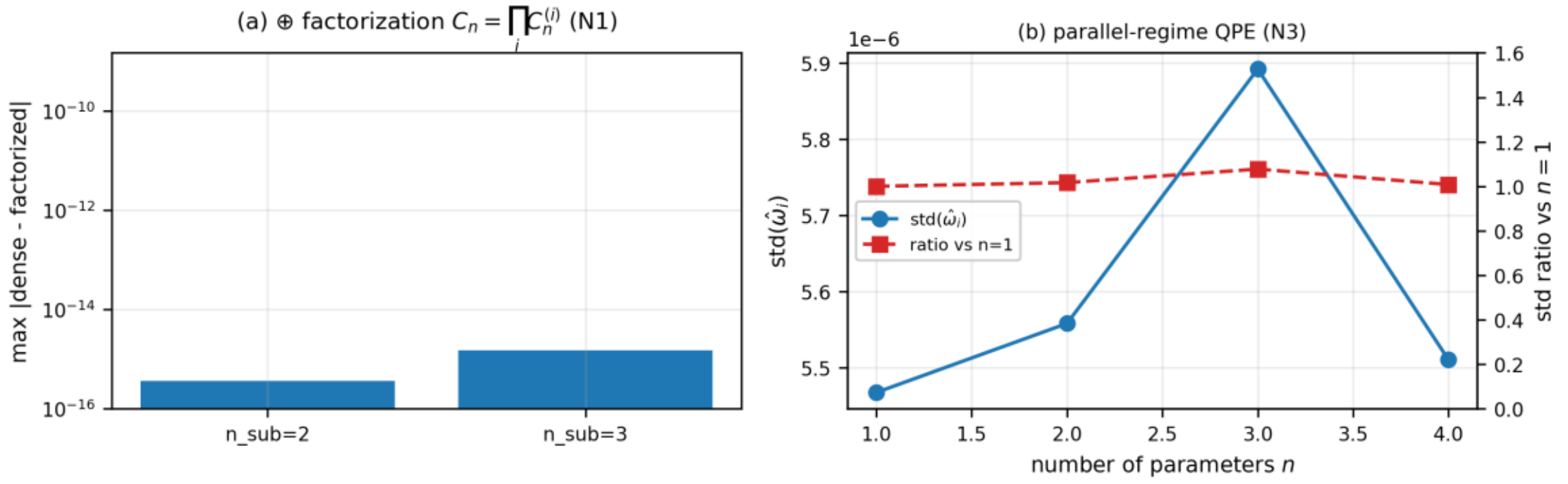

*Figure 12. Lower-bound numerics (suite N1–N3; SM §H). (a) ⊕ factorization of $C_n$ into per-subsystem factors (dense vs spectral vs product, machine precision); (b) the parallel-regime tensor-product QPE: estimation std and its ratio to $n = 1$ stay flat across $n = 1..4$ ($\gamma t = 5$, seed 20261022, $M = 20000$, 200 experiments).*

**Table 13. Lower-bound numerics** (slope constants: windowed $\gamma \cdot \max|dR_{\text{win}}/d\omega_1|$, Lorentzian $\gamma^2 \cdot \max|dR_L/d\omega_1|$).

| item | value |
|---|---|
| N1 factorization max diff (4 combos) | 1.49e-15 |
| N2 Lorentz slope $\gamma^2 \max|dR_L/d\omega_1|$ | 0.649519053 (analytic 0.649519053) |
| N2 window slope $\gamma \max|dR_{\text{win}}/d\omega_1|$ | 1.5707963 (analytic 1.5707963) |
| N3 std by $n = 1..4$ | 5.47e-06 / 5.56e-06 / 5.89e-06 / 5.51e-06 |
| N3 ratio vs $n = 1$ | 1.000 / 1.017 / 1.078 / 1.008 |
| N3 MC setting | seed 20261022, $M = 20000$ repetitions, 200 experiments, $t\gamma = 5$; std statistic (per-parameter; constant-success-probability statements add the $O(\ln n)$ union bound, Theorem 11) |

**Theorem 8 (resolution).** Any quantum algorithm with oracle access to controlled $e^{-iHt}$ in the oracle/probe and resource model of SM §E.1 (arbitrary classical adaptive $t$; free probe preparation; ancillas allowed; the fast-forwarding obstruction [26] is the physical backdrop; conversions to block-encoding calibers are governed by the oracle-conversion lemma of E.1 and never assumed free in the reverse direction) distinguishing $H^{(\omega_1)} = \omega_1 |1\rangle\langle 1|$ from $H^{(\omega_1+2\gamma)}$ with success $\geq 2/3$ needs total evolution time $T_{\text{total}} = \Omega(1/\gamma)$. Proof: the discrimination bridge of Lemma E1 with the two-level instance family (SM §E.2–E.4).

**Theorem 9 (windowed value task).** For the cosine-ramp window family (max-slope constant $c_2 = \pi/2$ in units $1/\gamma$; any window class with $c_2 = \Theta(1)$ gives the same bound with constants scaled by $c_2$ — $c_2$ is the shape parameter of the class), estimating $R_{\text{win}}$ at the max-slope point to additive $\varepsilon$ needs $T_{\text{total}} = \Omega(1/(\gamma\varepsilon))$ — a *product*, not a sum. The conjectured additive form $\Omega(1/\gamma + 1/\varepsilon)$ holds only as a task decomposition (parameter task vs readout task, different instance families and resource calibers). Matching upper bound: GQPE fixed-point physical time $D\Omega\pi/(W\gamma\varepsilon)$ — tight in both $\gamma$ and $\varepsilon$ dimensions.

**Lemma 10 (⊕ factorization).** For Kronecker-sum systems $H = \sum_i H_i \otimes \mathbb{1}_{\bar{\imath}}$, $D = \bigotimes_i D_i$, $|\psi_0\rangle = \bigotimes_i |0\rangle_i$: $C_n(t_n, \dots, t_1) = \prod_i C_n^{(i)}(t_n, \dots, t_1)$ — *time-domain* factorization (verified to $1.5 \times 10^{-15}$

across subsystem counts and orders); the frequency-domain resolvent does **not** factorize, $f(\sum_i H_i) \neq \otimes_i f(H_i)$ — the $\otimes \to \oplus$ correction that rescues the "factorized response" construction.

**Theorem 11 (parallel regime — an observation on the toy parameter family).** For $n$-parameter location on the toy family $H = \sum_i \omega_i g_i$ ($g_i = |1\rangle_i\langle 1|$, Kronecker sum; each $\omega_i$ to $\pm\gamma$): $T_{\text{total}} = \Omega(1/\gamma)$ (and the naive Σ-type $\Omega(n/\gamma)$ bound one might expect in the time caliber **does not hold**) — independent parameters accumulate phase in the *same* evolution segments (Lemma E1 per parameter, non-additive); a tensor-product QPE protocol achieves all $n$ parameters at joint constant success probability with $O((1/\gamma)\ln(n/\delta))$ — $\widetilde{\Theta}(1/\gamma)$ tight up to the logarithmic joint-success factor — an upper protocol for this parameter task only, not for the molecular response task (scope in SM §E.7; numerics: per-parameter std 5.5–5.9 × $10^{-6}$ flat in $n = 1..4$, ratios 1.000/1.017/1.078/1.008 — a standard-deviation statistic, Table 13, Fig. 12; constant-*probability* statements require the $O(\ln n)$ union-bound amplification, absorbed in $\tilde{O}$). Multi-parameter metrology precedents are variance-caliber results and cannot be read as query-cost sums [23].

**Theorem 12 (multi-variable depth, within the direct correlation-sampling class).** *Task:* estimate the windowed multi-variable response $R_n(\vec{\omega})$ at query points $\vec{\omega}$ to additive precision $\varepsilon$ (success $\geq 2/3$), at a resolution that separates a $2\gamma$ shift in any single frequency argument. *Class (direct correlation sampling):* a circuit together with its readout satisfying — (1) a serial circuit on a single **system** register, alternating $n+1$ dipole insertions with $n$ evolution **batches that partition the segments** (each evolution segment belongs to exactly one batch); (2) the batch-$j$ segment durations — and the time-register preparations realizing them coherently — are **query-independent**, fixed by the public parameters $(\gamma, \varepsilon, W)$ and depending on no query argument (classical choices or coherent superpositions; GQPE's per-variable time registers are the coherent realization); (3) the **system probe** and all operations on the system register are query-point-independent; (4) the readout is the **direct window transform in phase form** — the estimated value is a linear functional of the measured correlation values with *phase-type* weights,

$$F_{\vec{\omega}} = \sum_{\lambda} a_\lambda \, e^{i\sum_j \omega_j t_{\lambda,j}} \, \hat{C}(t_\lambda), \qquad \sum_{\lambda} |a_\lambda| \leq W_{\text{tot}},$$

where the amplitudes $a_\lambda$ are query-point- and instance-independent (a uniform-precision boundedness property), $t_{\lambda,j}$ are the (query-independent, clause 2) batch-$j$ times, and **the phase weights are the only query-point dependence of the whole procedure** (GQPE's inverse-QFT readout realizes exactly this structure: a fixed time-register window, with the frequency dependence carried by the QFT phases); the asymptotic statement below is made under a **uniform normalization convention** (its scope, stated positively: the convention covers normalized readouts — output distributions and normalized window series, GQPE's own objects; the *direct trapezoid transform* of Theorem 6(b)/Lemma D.4, with $\sum_\lambda |a_\lambda| = (\Delta t \sum w)^n \sim T^n$, lies *outside* this convention — and its $\gamma^{-(2n+1)}$ cost is exactly what the convention-carried bound would predict for it, so the class boundary is consistent rather than exclusionary): the weight–signal product $W_{\text{tot}}\mathcal{M}$ — the clause-4 weight bound times the instance signal scale $\mathcal{M}$ of E.8 ($\|C_n\|_\infty \leq \mathcal{M}$) — is uniformly bounded across the class; the readout may be the classical weighted sum or its **coherent (quantum) realization** — a measurement whose expectation equals $F_{\vec{\omega}}$; **no parametric re-modeling or extrapolation** is allowed (learn-then-assemble, sparse-recovery post-processing, and query-adapted time grids are thereby outside the class, registered in the open problem below). Within this class, under the uniform normalization convention above, any circuit-plus-estimator solving the task has serial depth $\Omega(n/\gamma)$, matched by GQPE's per-circuit $D/\eta_s = nW/(\pi\gamma)$ in both $n$ and $\gamma$: **GQPE is depth-optimal within the class under the stated normalization convention** — a restricted-class, convention-conditional statement (see the open problem below); constants and $W$, $\Omega$ factors aside — made explicit in E.8: the proof yields the normalization-carrying per-batch bound $T_j' = \Omega((c_0 - \varepsilon)/(\gamma W_{\mathrm{tot}}\mathcal{M}))$, with $c_0$ the straddle constant of E.8(i) and $\mathcal{M}$ the instance signal scale of E.8 ($\|C_n\|_\infty \le \mathcal{M}$) — the chain carries $W_{\mathrm{tot}}(\mathcal{M} + \varepsilon)$, and $\varepsilon \ll \mathcal{M}$ in the nontrivial regime — from which the advertised $1/\gamma$ scaling follows precisely under that convention. Proof in SM §E.8. *Scope:* this is a deliberate restriction, not a claim about arbitrary protocols — see the open problem below.

**Open problem.** Theorem 12 says nothing about protocols outside its class. Whether an arbitrary protocol can reach depth $o(n/\gamma)$ is open: parameter-location schemes followed by classical assembly, entangled probes, other oracle encodings, query-adapted grids, or circuits whose clock times serve several arguments at once. The question is nontrivial in both directions. Theorem 11 shows that a shared clock can accumulate the phases of all parameters at once, which is information-theoretically possible for the parameter task; on the other hand, the per-argument clock structure that our bound exploits is exactly what such protocols would have to circumvent.

**Lemma 13 (readout).** Estimating an expectation to additive $\varepsilon$ needs $\Omega(1/\varepsilon)$ controlled-$A$ calls (AE achieves it; classical sampling $\Omega(1/\varepsilon^2)$) — circuit-repetition caliber, orthogonal to $T_{\mathrm{total}}$; both are reported separately.

**Corollary 14 (the $\gamma^n$ verdict, by caliber).** The status of the $\gamma$-dependence splits into resource calibers, each either closed or explicitly open:

1. **Parameter-location subtask** (locate each of $n$ independent frequencies to $\pm\gamma$): total evolution time $\widetilde{\Theta}(1/\gamma)$ — tight in both directions up to the $O(\ln n)$ joint-success factor (Theorem 11 lower bound; tensor-product QPE upper bound). The $n$-parameter bound is *not* $\Omega(n/\gamma)$ in this caliber (the parallel bonus).
2. **Windowed single-point value task** (estimate $R_{\mathrm{win}}$ at a query point to additive $\varepsilon$): total evolution time $\Theta(1/(\gamma\varepsilon))$ — tight in both $\gamma$ and $\varepsilon$ (Theorem 9 lower bound; GQPE fixed-point upper bound, *in the total-time caliber including AE repetitions*: $M = \Theta(1/\varepsilon)$ controlled calls, each with evolution $\Theta(1/\gamma)$, give $\Theta(1/(\gamma\varepsilon))$ — the same type as the lower bound; this inclusion is what makes the match hold).
3. **Full-spectrum task**: per-circuit resolution $\gamma^{-1}$ is achievable, by GQPE output-distribution sampling, and that is a single-circuit statement. In the single-circuit-resolution caliber, $\gamma^{-n}$ is the price of $n$-dimensional grid *coverage*, not of single-point information. In the total-cost caliber the sample-side question is open: whether $\Omega^4\varepsilon^{-4}$ distribution sampling is necessary has not been settled, so no end-to-end "necessary and sufficient" claim is made at this level. The multi-variable $n$-factor is a depth phenomenon, $\Omega(n/\gamma)$ **within the direct correlation-sampling class** (Theorem 12), and open for arbitrary protocols. A fourth entry: **Lorentzian object, additive precision,** $s^* = \Theta(1)$ **line-strength families** — $\Omega(1/(\varepsilon\gamma^2))$ per point with exponent-matching upper bounds across the stated oracle models (Corollary 15).

With (1)–(4), the suspicion of Kharazi et al. that "$\gamma^n$ may be a dense-grid artifact" is confirmed in the parameter/coverage sense (single-circuit-resolution caliber), with matching upper bounds supplied by [10] in calibers (2)–(3); the total-cost caliber remains open on the sample side.

---

# 7 The six-route unified comparison

## 7.1 The table

**Table 14a. Per-route normalized single-point cost map: each route at relative-to-peak precision $\varepsilon_{\rm rel}$ on its own measured object** ($U_H$ queries; the normalization scales differ per route and are defined in the table — **a normalized cost map, not a same-physical-task ranking**; the measured-object column states the conversion status to the physical target; route ③ is excluded as a cited, differently-normalized estimate, see Table 14b; the line strength $s$ and the sparsity of Sec. 5.3 are distinct quantities, the latter appearing only in Remark 7.1).

| route | normalization scale of the measured object | measured object / conversion to physical target | cost at relative-to-peak $\varepsilon_{\rm rel}$ |
|---|---|---|---|
| ① indicator binning (tight-$\zeta$ reading, Lemma Q) | $s^*$ (line strength captured by a resolution bin) | bin intensity (integral semantics); to the Lorentzian density needs window/area conventions | $\tilde{O}\big((\alpha/\gamma)(\beta^2/s^*)\,\varepsilon_{\rm rel}^{-1}\big)$; published accounting carries the extra $\Theta(\alpha+1+\omega)$ (Sec. 4.4) |
| ② resolvent (Thm 1; $D'=D$) | $\tilde{\mu}_2(\omega)\approx c\,s^*$ | $\bar{S}_\gamma=\gamma\,{\rm Im}R$ — exact multiple of the physical density (conversion exact, Sec. 4.1 dictionary) | $\tilde{O}\big(((\alpha+\omega)/\gamma)(\beta^2/\tilde{\mu}_2)\,\varepsilon_{\rm rel}^{-1}\big)$; $\sqrt{g_2}$-AE readout: $((\alpha+\omega)/\gamma)(\beta/\sqrt{\tilde{\mu}_2})\,\varepsilon_{\rm rel}^{-1}$; exact lineshape + disjoint budget |
| ④ time-domain direct (Thm 6(a)) | $S_{\rm corr}:=\mathcal{M}=\sup\lVert C_n\rVert$ (**correlation scale, not a spectral peak**) | time-domain correlation $C_n$; conversion to a frequency-domain peak depends on window, damping, and lineshape (not automatic) | $\tilde{O}\big(\alpha\beta^{n+1}T_{\rm tot}/\mathcal{M}\cdot\varepsilon_{\rm rel}^{-1}\big)$; $T_{\rm tot}$ is the task's evolution time |
| ⑤ GQPE fixed point | $\tau s$ (windowed peak of the scaled response; $\tau=\pi/W$) | normalized windowed response $R_s=R/\tau$ (multiply by $\tau$; Sec. 5.2) | $\Theta(W\beta^{n+1}/(\gamma\,s\,\varepsilon_{\rm rel}))$; no $\alpha$ only in the additive-on-$R$ caliber (Sec. 5.2) |

**Table 14b. Native-semantics reference map of the six routes for the $n$-th order response ($U_H$ queries; $N_e$ = electron count; $W = \omega_{\max}$, centered encoding $\alpha = W/2$ / uncentered $\alpha \geq W$; entries are quoted in each route's native precision semantics with the conversions displayed — additive-on-$R$ vs relative-to-peak differ by the peak scale (Sec. 5.2) — and are** not mutually comparable**; use Table 14a for the unified single-point comparison; the status column separates proved from extrapolated entries; Õ suppresses all logs).**

| route | single-point ($\tilde{O}$, $U_H$) | full-spectrum strategy ($\gamma$-power × rest) | subnorm. | readout | status |
|---|---|---|---|---|---|
| ① indicator binning [1] | as published: $n\alpha^{n+1}\beta^{n+1}/(\gamma\varepsilon)$ (nested single bin; degrees *add*); tight-$\zeta$ reading (Lemma Q, ours): $n\alpha\beta^{n+1}/(\gamma \cdot \text{signal} \cdot \varepsilon)$ — same form as ② | adaptive bisection $\gamma^{-n}$ (Eq. 33: $\alpha^{2n}\beta^{n+1}/\gamma^n\varepsilon$; the tight-$\zeta$ reading removes the subnormalization's $\alpha^n$: one $\alpha$ at $n = 1$); per-grid $\gamma^{-(n+1)}$ | $\zeta_\omega^{(n)} \sim (3\alpha)^n\beta^{n+1}$ (published caliber; order-of-magnitude $n$-order extrapolation — the $n = 1$ exact factor is $\alpha + 1 + \omega$, Sec. 4.4) / $\beta^{n+1}$ (tight, Lemma Q) | AE $\varepsilon^{-1}$ | cited theorem [1] + our tight-caliber reading (Sec. 4.4) |
| ② resolvent (this work, Thm 1/4) | $n = 1$: $\tilde{O}(((\alpha + \omega)/\gamma)(\beta\beta'/\tilde{\mu}_2)\varepsilon_{\text{rel}}^{-1})$; tight $\zeta = \beta\beta'$ (Lemma Q); $n = 2$ (Thm 4) explicit, $n \geq 3$ structured | per-frequency sampling; strong for continuous spectra (Sec. 4.7) | $\beta\beta'$ (tight) / $\xi^2\beta^{n-1}$ (prep.) | AE $\varepsilon^{-1}$ | **this work**: Thm 1 proved; Thm 4 proved ($n = 2$); $n \geq 3$ by induction (App. B.3) |
| ③ linear system [2] modernized | $N^7 N_e/(\gamma\varepsilon^2)$ $(n = 1)$ [b] | solve per $\omega$ | $\kappa(A) \approx \lVert H \rVert/\gamma$ | QSP solver + expectation estimate $O(N^4/\varepsilon^2)$ (not AE) [b] | estimate quoted from [2] Discussion; layer split ours (fn. b) |

| route | single-point ($\tilde{O}$, $U_H$) | full-spectrum strategy ($\gamma$-power × rest) | subnorm. | readout | status |
|---|---|---|---|---|---|
| ④ time-domain AE (Thm 6; direct correlation sampling) | Thm 6(a): $\alpha\beta^{n+1}T_{\mathrm{tot}}/\varepsilon$ | per-grid $\gamma^{-(2n+1)}$ (worst-case noise, Lem. D.4) / distribution sampling $\gamma^{-1}$ ($\Omega^4\varepsilon^{-4}$) / sparse recovery (fn. c) | $\zeta_t = \beta^{n+1}$ (no indicator); GQPE $\Omega = \beta^{n+1}$ | AE $\varepsilon^{-1}$ | Thm 6 proved (all $n$; sparse recovery: Prop. 7 stated at $n = 1$ with stability quoted from the classical literature (App. F); ANM branch conditional (Remark 7.1); $\beta^{n+1}$ is the $n$-order extrapolation (fn. c) |
| ⑤ GQPE multi-phase [10] | $(\pi/2)n\beta^{n+1}/(\gamma\varepsilon)$ with $\varepsilon$ additive on $R$ (Sec. 5.2); relative-to-peak: $\Theta(W\beta^{n+1}/(\gamma\, s\, \varepsilon_{\mathrm{rel}}))$ (the GQPE fixed point lives here, not in route ④) | output-distribution sampling $\gamma^{-1}$ (samples $\Omega^4/\varepsilon^4$; fixed-point full-spectrum prefactor $(\pi^4/16\alpha^3)\, n\, \beta^{4(n+1)}/(\gamma\varepsilon^4)$; complete-square $(\pi^2/4\alpha)\, n\, \beta^{2(n+1)}/(\gamma\varepsilon^2)$) | $\Omega = \prod \lVert V^{(j)} \rVert_1 = \beta^{n+1}$ | AE $\varepsilon^{-1}$ | cited theorem [10] + our conversion (Sec. 5.2) |
| ⑥ ancilla-free shot [8, 9] | — (no complexity theorem) | fixed-grid shot | — | shot $\varepsilon^{-2}$ | no complexity theorem (mechanism descriptions only) |

[b] $\varepsilon^2$**-caliber mechanism (carried with this table by construction).** Route ③'s $\varepsilon^{-2}$ is *not* a same-task one-power deficit against routes ①②④⑤: modern QSP linear-system solvers carry the solver residual only in logarithms (suppressed by $\tilde{O}$); the $\varepsilon^{-2}$ sits in an *expectation-estimation layer* — $A^\dagger A$ has $O(N^4)$ terms and the solution observable is estimated by sampling at $O(N^4/\varepsilon^2)$ cost — a layer the AE-readout routes replace by amplitude estimation ($\varepsilon^{-1}$, Lemma 13). Cross-table comparisons must note the mechanism difference (measurement-model layer vs solver layer); the original Cai et al. HHL form $O(\log(D)s^2\kappa^2/\varepsilon)$ does not separate the measurement-side $\varepsilon$ scaling. [c] Route ④'s sparse-recovery entry: ESPRIT (Proposition 7, stability quoted from the classical literature) costs $\tilde{O}((W/\Delta_{\text{sep}})\,\alpha\beta^2/(\gamma\varepsilon_q))$ at $n=1$; the conditional ANM branch (Remark 7.1) would cost $\tilde{O}(s\,\text{polylog}\cdot\alpha\beta^{n+1}/(\gamma\varepsilon_q))$; the displayed $\beta^{n+1}$ is the $n$-order extrapolation through Theorem 6(a)'s subnormalization $\zeta_t=\beta^{n+1}$.

## 7.2 Lower-bound matching and the $\gamma$-power genealogy

**Table 15. Lower bounds vs matching upper bounds (Sec. 6).**

| task / caliber | lower bound | matching upper bound | status |
|---|---|---|---|
| parameter location ($n$ independent frequencies; *time* caliber) | $\Omega(1/\gamma)$ — parallel regime, not Σ-type $n/\gamma$ | tensor-product QPE $O(\ln(n/\delta)/\gamma)$ | tight up to joint-success logarithm (Thm 11) |
| windowed value (single point) | $\Omega(1/(\gamma\varepsilon))$ (product) | GQPE fixed-point physical time $D\Omega\pi/(W\gamma\varepsilon)$, total-time caliber incl. AE repetitions | tight in $\gamma,\varepsilon$ (Thm 9) |
| multi-variable query-point response (*depth* caliber; **restricted class**: direct correlation-sampling, phase-only readout, **normalization-dependent**) | $\Omega(n/\gamma)$ | GQPE per-circuit $nW/(\pi\gamma)$ | tight in $n,\gamma$ **within the restricted class under the stated normalization convention** (Thm 12); arbitrary protocols open |
| readout | $\Omega(1/\varepsilon)$ | AE | optimal (Lem 13) |
| Lorentzian single point, additive $\varepsilon$ ($s^*=\Theta(1)$ families) | $\Omega(1/(\varepsilon\gamma^2))$ (E.1 total-time caliber) | resolvent route $\tilde{O}(\alpha\beta^2/(\varepsilon\gamma^2))$ ($U_H$ caliber, Thm 1) | $\gamma,\varepsilon$ exponents matched across the stated oracle models (Cor 15) |

| task / caliber | lower bound | matching upper bound | status |
|---|---|---|---|
| full spectrum (per-circuit resolution) | $\gamma^{-1}$ necessary for the parameter subtask $\Rightarrow$ $\gamma^{-n}$ is a coverage artifact | GQPE distribution sampling $\gamma^{-1}$ per circuit | per-circuit achievability settled (upper bound); per-circuit depth lower bound only within the direct class (Thm 12); sample side $\Omega^4\varepsilon^{-4}$ open (Cor 14) |

**$\gamma$-power genealogy** (full-spectrum task; caliber labeled per entry and $\varepsilon$-semantics where it differs — single-circuit depth and total cost are never mixed): per-grid binning (no bisection) $n+1$ (total); adaptive bisection $n$ (total); per-frequency grid of route ② $n+1$ (**family F1**: total $(W/\gamma)$ coverage × per-point $\gamma^{-1}$ at fixed strengths; the per-point exponent is $\varepsilon$-semantics-dependent — Sec. 6 dictionary); per-grid time domain $2n+1$ (total, worst-case sampled noise, Lemma D.4, family F2; $2n$ in the fixed-strength family F1; $3n/2+1$ under Proposition 7's independence assumption); GQPE output-distribution sampling 1 (**single-circuit** resolution; samples $\Omega^4/\varepsilon^4$, sample side open); single-ancilla random-$t$ $2n+1$ (total, with the $(W/\pi)^{2n-2}$ penalty); sparse recovery — ESPRIT 2 (total, $\tilde{O}((W/\Delta_{\text{sep}})\,\alpha\beta^2/(\gamma\varepsilon_q))$; **conditional** quantum sampling-cost accounting under the quoted classical stability assumptions), ANM variant 1 with poly($s$)log points *(conditional — Remark 7.1, random-signs hypothesis unsatisfied)*. $\gamma^{-n}$ is never the cost of a single-point estimate; it is the price of covering an $n$-dimensional grid — the nested single-bin degree is *additive* $(n\cdot\alpha/\gamma)$, and the $(\alpha/\gamma)^n$ of Eq. 33 enters only through bisection coverage.

### 7.3 The $n=2$ check (three accounts)

*Cost account:* time-domain per-grid $n=2$ full spectrum $2(W/\pi)^2\alpha\beta^3L^5/(\gamma^5\varepsilon A_s)$ (worst-case noise accounting, family F2; $L$ the window logarithm of D.2) vs Eq. 33's $\alpha^4\beta^3/(\gamma^2\varepsilon)$ — the per-grid time-domain upper bound exceeds the bisection upper bound throughout the practical range (worst-case noise propagation multiplies the naive-accounting grid/bisection ratio by $(L/\gamma)^n$; Sec. 5.1); distribution sampling reaches $\gamma^{-1}$ per circuit (sample side open), and the sparse-recovery entry with conditional quantum sampling-cost accounting (under the quoted classical stability assumptions) is ESPRIT at $\gamma^{-2}$ (the $\gamma^{-1}$ ANM entry is conditional, Remark 7.1). *Structure account:* $\oplus$ factorization holds in the time domain (numerically $1.5\times10^{-15}$) and fails in the frequency domain; tensor-product 2D line spectra recover supports 20/20 (2D-ESPRIT). *Symmetry account:* the C2v selection-rule zero ($7.65\times10^{-13}$) and the $\kappa_n$ framework (SM §B.1) both reappear on the $n=2$ objects. **Verdict:** structurally no new obstacle appears at $n=2$ (factorization and support bookkeeping carry over; SM §B.6 closes the object-level identity); the end-to-end bill for the physical $\beta_{ijk}$ is governed by the component-grade coherence $\kappa_{\text{asm}}$ (SM §B.6), extreme cancellation being an instance-dependent cost factor; for $n\geq3$ the structured statement (time-domain factorization for all $n$; none in the frequency domain) is the correct form.

## 8 The empirical phase diagram: which spectrum, which arm

Three measurement arms were compared on common calibers (Fig. 13; medians in Tables 11–12). **Scope.** All conclusions in this section apply to the tested ensembles only: (i) a 6-line Lorentzianized test spectrum ($\gamma = 0.31818$, $n_{\text{grid}} = 512$, bins {16,32,64}, $\varepsilon_{\text{rel}} \in \{0.1, 0.03, 0.01\}$, 7 seeds × 9 configurations), where line generation enforces separation $\geq 2\gamma$ so that the ESPRIT arm is tested inside the separation hypothesis of Proposition 7; (ii) 2D tensor-product line spectra (20 seeds); (iii) continuous synthetic densities (E2). The dense-spectrum time-domain cell is untested and marked "open" in Fig. 13 — an untested regime, not a negative result. The arms run at matched $\varepsilon_{\text{rel}}$ with their resource columns reported separately (Table 7), so the comparison is at unmatched total $U_H$ resources; it is not a preregistered head-to-head, and the operating rule below applies to the tested ensembles. The three arms are the bin arm (disjoint indicator histograms), the kernel arm (Lorentzian or Gaussian convolution sampling, route ②), and the time-domain arm (damped-sinusoid sampling followed by ESPRIT, route ④ with the classical stage of Proposition 7).

- **Simple (Lorentzianized, few-line, well-separated) spectra → kernel arm.** On that spectrum: kernel median $L^2$ $3.2 \times 10^{-4}$–$3.8 \times 10^{-3}$; bin 0.057–0.229; ESPRIT 0.199–0.792. The bin arm's $L^2$ is dominated by the *staircase discretization bias* of 16/32/64-bin histograms against the 512-point smooth target. That is a resolution mismatch rather than an algorithmic penalty: bin error improves only weakly and sublinearly along the $\varepsilon_{\text{rel}} = 0.1 \to 0.03 \to 0.01$ steps, with median per-step factors 1.03–1.94 from Table 12. The kernel/bin ratio, suppressed in all 63 rows with a minimum of 31.5 ×, should therefore be read as a statement about resolution-matched versus staircase readout, not as an algorithm-versus-algorithm verdict.
- **Sharp off-grid lines and 2D joint sparsity:** for the tested ensembles of this type, the time-domain arm provides grid-independent localization (its most favorable tested regime; no competing arm was run there). ESPRIT's off-grid peak accuracy is grid-independent ($\text{err}/\varepsilon_q \approx 0.45$ plateau); 2D tensor-product supports recover 20/20. But on the simple-spectrum regime its spectral $L^2$ *loses* to kernel inversion — in-domain (the separation hypothesis holds there), a boundary of validity rather than a violation of the proposition; no competing arm was run in the sharp-off-grid ensemble.
- **Continuous/dense spectra → kernel arm (with the deconvolution rates of Sec. 4.7).** Kernel-vs-bin at equal resolution: 2.8 × cheaper in degrees and 7 × more accurate; bin noise grows with bin count (per-bin SNR mechanism: $n_{\text{bins}}$ unknowns each constrained by exactly one measurement — no noise averaging — while kernel rows form a redundant frame).
- **Bin + bisection retains one role:** initial screening / peak finding (flat-top semantics; Sec. 4.6 criterion 3).

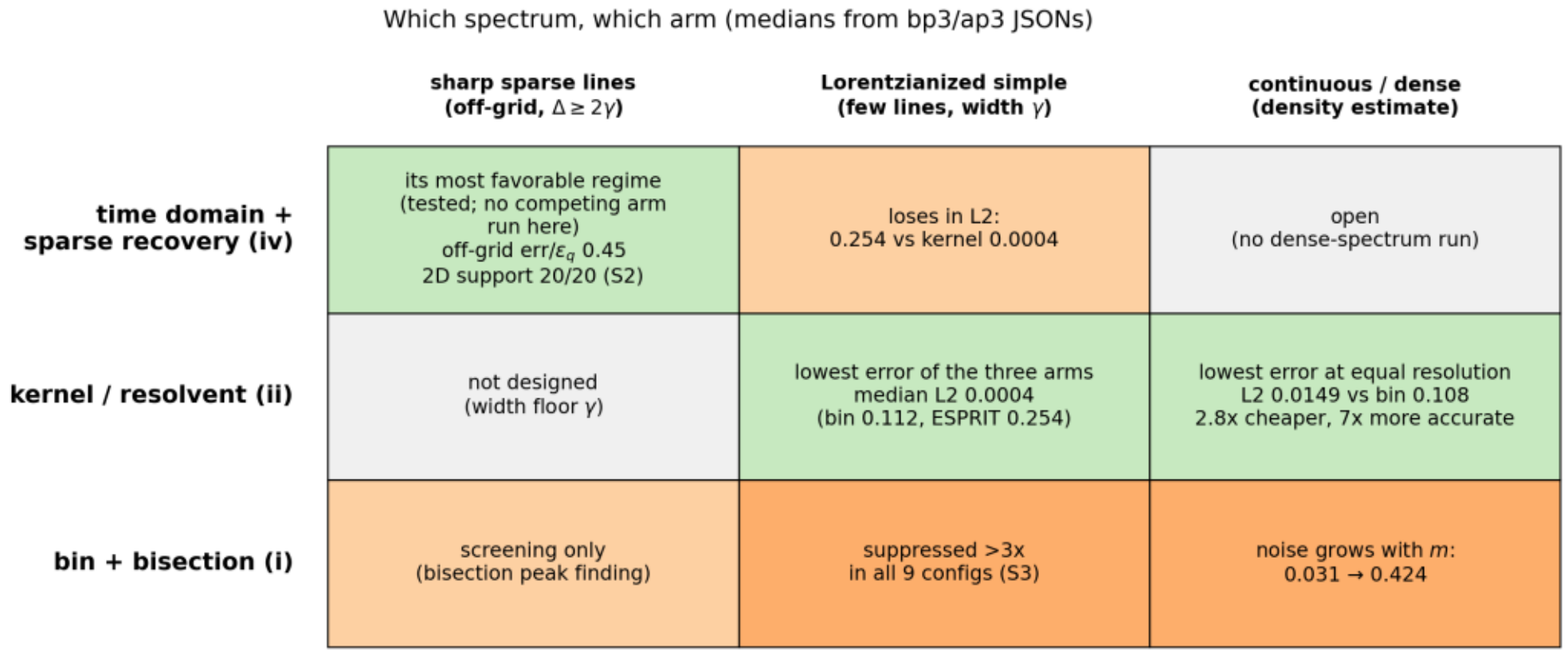

*Figure 13. The empirical phase diagram ("which spectrum, which arm"). Cell entries are medians computed by the figure script from the archived suite results (S1/S2/S3, E2); "open" marks an untested regime (time-domain arm on dense spectra) rather than a negative result.*

The resulting heuristic operating rule — *peaks: bin + bisection (erf); lineshape and continuous spectra: kernel (Lorentz/Gauss); sharp off-grid or joint 2D sparsity: time-domain + sparse recovery* — is a heuristic supported by the tested ensembles above; it is not a theorem, and broader preregistered comparisons at matched total resources would be required to elevate it beyond the tested scope.

---

## 9 Scope and open problems

**What the results rest on.** Theorem 4 and Lemma 5 use the D1 (uniform-$\gamma$) convention. The D1 $1/(i\gamma)^2$ static-channel singularity is quantified ($B_{zzz} \sim 6d_{00}^3/\omega^2$ to 0.51%); the Orr–Ward D2/D3 bookkeeping with state-dependent $\Gamma_m$ would need per-state kernels (degree $\max_m \Gamma_m^{-1}$) and is not treated here. Assumption A1 (exact ground state) and the uniform phenomenological $\gamma$ are inherited from the reference setting; temperature (A4) and beyond-dipole couplings (A3) lie outside the model. Proposition 7 carries three assumption-grade inputs — independence of the AE errors (structural argument, no counterexample), the ANM sample-count law (cited, not numerically verified), and the Bhaskar et al. denoising constants and poly(SRF) powers (quoted at skeleton level, SM §F). The readout-layer improvement requires a positive-semidefinite sandwich and a known-or-doubled signal scale (Remark after the Sec. 4.3 corollary); the $n = 2$ physical assembly carries the component-grade coherence $\kappa_{\text{asm}}$, which is not certifiable a priori (SM §B.6), with near-zero components handled by absolute noise floors; the $A_s$ bookkeeping splits into the (F1) fixed-strength and (F2) fixed-target-scale instance families (Thm 6(b), D.2) and comparisons must declare which is in use; GQPE's "no $\alpha$" holds in the additive-precision caliber on $R$ (Sec. 5.2).

**What is settled.** The $\Sigma$-type $n/\gamma$ expectation does not hold in the time caliber (Theorem 11). The $\otimes \to \oplus$ correction is applied throughout (Lemma 10). The Lorentzian object's $\gamma$- and $\varepsilon$-exponents are matched across the stated oracle models (Corollary 15). In the single-circuit-resolution caliber, $\gamma^{-n}$ is a coverage artifact.

**What remains open.** (i) Sample-side optimality of output-distribution sampling ($\Omega^4 \varepsilon^{-4}$): whether $\Omega^4$ is required, and whether sparse priors beat it at matched output error and confidence (the sparse-recovery arm's random-subset caliber is one candidate; the comparison is same-caliber on the query side, while the classical post-processing accounts differ). (ii) The bandwidth-decoupled ANM sample count under nonnegative strengths. (iii) Independence of the AE errors in Proposition 7. (iv) Extreme-cancellation regimes of $\kappa_{\text{asm}}$. (v) The D2/D3 damping bookkeeping. (vi) The pre-asymptotic deconvolution domain: the $1/\gamma$ asymptote is not reached in the tested $\varepsilon$ range (envelope [16.9,40.7] vs asymptote 100), so smaller-$\varepsilon$ validation and mixed peak+continuous regularization layering remain to be done. (vii) Degree extrapolation to $\tilde{\gamma} \sim 10^{-4}$ ($d \sim 10^5$) and peak-to-peak dense-line resolution, beyond the validated range ($\tilde{\gamma} \in \{0.02, 0.01, 0.005\}$, 30 cells, pyqsp chain to $d = 1830$). (viii) A matching upper bound for the parameter and windowed tasks *within a single oracle model*: the present matching is exponent-level across models (Table 15). (ix) Whether arbitrary protocols can break the $n \cdot \gamma^{-1}$ depth of Theorem 12; the bound is proved inside the restricted direct correlation-sampling class.

---

## 10 Conclusion

We have compared six algorithmic routes for quantum response-function estimation within one framework, with the conversions between their resource calibers stated explicitly.

On the resolvent route, exact lineshapes cost the same per point as indicator binning once the removable $\Theta(\alpha + 1 + \omega)$ factor in the published binning accounting is taken out. That correction benefits both routes, and its definitional ratio is $\alpha + 1 + \omega$. The same route comes with an error budget that covers both finite-$\gamma$ boundaries without double counting, a frequency-independent preparation circuit, and a square-root-kernel readout that improves the $\beta^2/\tilde{\mu}_2$ layer to $\beta/\sqrt{\tilde{\mu}_2}$ where the sandwich is positive.

The time-domain route feeds a sparse-recovery arm. The stability of that arm is imported from the classical literature (Proposition 7, under a stated perturbation-regime hypothesis).

The lower bounds settle the parameter and windowed single-point calibers, $\widetilde{\Theta}(1/\gamma)$ and $\Theta(1/(\gamma\varepsilon))$. They also add a Lorentzian-object corollary, $\Omega(1/(\varepsilon\gamma^2))$ under additive precision for $s^* = \Theta(1)$ line-strength families, whose exponents match the upper bound across the stated oracle models. In the single-circuit-resolution caliber, $\gamma^{-n}$ is a coverage artifact. The multi-variable $n$-factor sits in serial depth inside the restricted direct correlation-sampling class, and readout is optimal at $\Omega(1/\varepsilon)$. The $n = 2$ assembly carries the component-grade coherence $\kappa_{\text{asm}}$. The phase diagram reports the tested ensembles.

Scope conditions and open problems are collected in Sec. 9.

## Data availability

All numerical data and code supporting the findings of this paper are openly available in the Gitee repository **https://gitee.com/xie-qingxing/quantum-response-functions** (data and code for "Quantum Algorithms for Molecular Response Functions"). The package contains: the nine archived numerical result files underlying every figure and table (Figs. 1–13, Tables 1–15); the nine suite scripts that generate them, plus two supporting files (a standalone propagation check and a read-only molecular-integration library); the deterministic figure/table generation script with its per-figure provenance manifest; the random seeds and SHA-256 integrity fingerprints of all archived files; and step-by-step reproduction instructions. The Supplemental Material with the full derivations (sections A–J) accompanies this manuscript.

## Funding

This work is supported in part by the National Natural Science Foundation of China (Grant No. 12504267), the Natural Science Foundation of Hubei Province (Grant No. 2025AFB274).

## Competing interests

The authors declare no competing interests.